\documentclass[twocolumn,showpacs,preprintnumbers,aps,prc,10pt,nofootinbib]{revtex4-1}

\usepackage[english]{babel} 	
\usepackage{graphicx}
\usepackage{amssymb} 			
\usepackage{amsmath} 			
\usepackage{booktabs}
\usepackage{tabularx}
\usepackage{braket}

\usepackage[cm]{fullpage}

\allowdisplaybreaks

\begin{document}


\title{Capture reactions into borromean two-proton systems at rp-waiting points}

\author{D. Hove, A.S. Jensen, H.O.U. Fynbo, N.T. Zinner, and D.V. Fedorov}
\affiliation{Department of Physics and Astronomy, Aarhus University, DK-8000 Aarhus C, Denmark} 

\author{E. Garrido}
\affiliation{Instituto de Estructura de la Materia, IEM-CSIC, Serrano 123, E-28006 Madrid, Spain}

\date{\today}


\begin{abstract}
We investigate even-even two-proton borromean systems at prominent intermediate heavy waiting points for the rapid proton capture process. The most likely single-particle levels are used to calculate three-body energy and structure as a function of proton-core resonance energy.  We establish a linear dependence between two- and three-body energies with the same slope, but the absolute value slightly dependent on partial wave structure. Using these relations we estimate low-lying excited states in the isotones following the critical waiting points. The capture rate for producing a borromean bound state is described based on a full three-body calculation for temperatures about $0.1-10$~GK. In addition, a simple rate expression, depending only on a single resonance state, is found to comply with the three-body calculation for temperatures between $0.1$ and $4$ GK. The rate calculations are valid for both direct and sequential capture paths. As a result the relevant path of the radiative capture reactions can be determined. We present results for $E1$ and $E2$ photon emission, and discuss occurrence preferences in general as well as relative sizes of these most likely processes. Finally, we present a method for estimating proton capture rates in the region around the critical waiting points.
\end{abstract}


\maketitle


\section{Introduction}

A number of proton dripline nuclei are of particular interest as
waiting points in the rapid-proton capture (rp) process expected to be active in the accretion of a close binary system containing a neutron star and resulting in an X-ray burst \cite{wal81,sch98,wal97,sch01,tho09}.  When the proton binding energy becomes negative at the dripline
another proton is needed to produce the borromean system (a bound system with unbound subsystems). The
effective lifetime of the critical waiting points in a stellar environment is a central quantity in the understanding of this astrophysical process. This depends crucially on both proton binding of the waiting point plus one and two protons, as well as the reaction rate forming these nuclei \cite{bro02,gor95}. The current estimates still result in an uncertainty in the effective lifetime of several orders of magnitude \cite{sch06,koi99}. The energy, capture time and capture mechanism can be explored through three-body calculations.  Relatively few full three-body results have been published, although many
capture rates have been estimated using various approximations \cite{gor95,bro02,gri05}. So far the three-body results have been limited to nuclei lighter than $^{38}\text{Ca}$ \cite{gri05,gor95}. However, three heavier critical waiting points exists, which have as of yet not been treated from a few-body perspective. These critical waiting points are $^{64}\text{Ge}$, $^{68}\text{Se}$, and $^{72}\text{Kr}$ \cite{sch98, woh04}. Recent efforts suggest $^{64}\text{Ge}$ is of less importance than previously thought, while $^{68}\text{Se}$, on the other hand, is thought to be of prime importance \cite{sch07,tu11}. 

Weakly bound nuclear states have been successfully described as
few-body structures in a number of cases for a long time \citep{jen04,zhu93}. The most thorough and
abundant theoretical investigations exist along the neutron dripline \cite{han87,bar06}, but also
excited states of ordinary nuclei have revealed this structure \cite{jen04}, and
recently a few medium-heavy alpha-dripline nuclei were suggested to be
of two-alpha + core structure \cite{hov14}. Few-body formalism is most often applied to the very light nuclei, where the constituent particles have a less intricate structure. For instance, few-body models have been applied successfully in describing the bridging of the mass gaps at $A= 5$ and $8$ \cite{efr96,fed10}. Likewise, three-body models have been applied for two-neutron plus core systems to provide astrophysically relevant production rates at the neutron dripline \cite{die10,bar06} again for the very light systems. The next step to the proton dripline nuclei and two-proton structure has been investigated for special, relatively light nuclei \cite{gio02,gar04b,gri05,gor95}, but much less for heavier systems. For the heavier nuclei the increased Coulomb interaction severely complicates at least the numerical calculations, if not the conceptual picture. 

The location of the proton dripline - which is rather irregular - is fairly well established at least up to around the medium-heavy nuclei \cite{erl12,orm97,bro02,tho04}, but details
of the nuclear properties are often very scarce.  The most important
quantity is of course the binding energy and the related stability which
provides the definition of the dripline. Near the neutron dripline it is
established that low-angular momentum single-particle states produce
spatially extended two-neutron halo nuclei \cite{jen04}. This few-body picture is
less effective at the proton dripline since the unavoidable Coulomb
interaction would produce a confining barrier for bound nuclei.
However, a cluster structure of decoupled core and proton degrees of
freedom still constitutes a fair description provided the core is relatively tightly bound, and the few-body binding energy is very small or perhaps even
negative.

Performing two and three-body calculations at the proton dripline
requires information about the single-proton orbits, in particular their
energy, as well as spin and parity relative to the core
nucleus.  Measured values are in most cases not available for these
ground and low-lying excited states, which should be occupied by the
protons just before or after reaching the dripline. Instead, trends from neighbouring nuclei and especially
the results from mean-field calculations are used. The inherent uncertainties can
be ascertained by variation of these input parameters.

The purpose of this paper is first of all to provide characterizing
information about the rp process at waiting points
for intermediate heavy nuclei along the proton dripline. The intend is to provide an initial method for approaching nuclei around the critical waiting points from a three-body perspective. The
borromean three-body structure is crucial as an intermediate structure
which therefore first must be investigated.  We shall determine
structure and constraints for the decisive two-body proton-core
energies. This is used to present a general method for estimating proton capture rates given very sparse experimental information. 

In Sec.~\ref{sec theo} we briefly give the
theoretical framework and pertinent formulae for both the few-body formalism and the three-body reaction rates, as well as the parametric choices
of the potentials. Section \ref{sec 3b grnd} contains a detailed
description of the ground state properties of the borromean proton
dripline systems of interest. Both the relation between two- and three-body energies and the structure of the ground state is examined in detail. Section \ref{sec pro} gives calculated results for the radiative capture reaction rates, based on the most likely two-body energies. This also includes a necessary examination of the most likely excited states. Section
\ref{sec impli} discusses limiting values of the basic physical
parameters, and predict borromean structure and the related
reaction mechanism at the astrophysical waiting points. This is combined to provide a method for estimating proton capture rates around the critical waiting points. Finally, we
briefly summarize and conclude in Sec.~\ref{sec con}.

\section{Theoretical framework \label{sec theo}}

Our aim is to calculate proton dripline three-body structures and
capture rates from continuum two-proton states into the corresponding
borromean ground state at medium heavy waiting points. For this we
need properties of ground and continuum states as well as
the derived rates and cross sections.  The general theoretical
background can be found in Refs.~\cite{nie01,gar04,die11,gar11}, but for completeness we
shall here collect the ingredients necessary to explain
notation, calculations and results.  We shall only discuss two-proton
three-body systems.

\subsection{Three body procedure \label{sec 3b}}

We outline briefly our method of hyperspheric adiabatic expansion of
the Faddeev equations in coordinate space \cite{nie01}.  The system of
two protons and a core can be described by three sets of the relative
Jacobi coordinates, $(\mathbf{x}_i, \mathbf{y}_i)$, defined as
\begin{align}
\mathbf{x}_i &= (\mathbf{r}_j - \mathbf{r}_k ) \sqrt{\frac{\mu_{jk}}{m}}, \\
\mathbf{y}_i &= \left(\mathbf{r}_i - \frac{m_j \mathbf{r}_j + m_k \mathbf{r}_k}
{m_j + m_k} \right) \sqrt{\frac{\mu_{jk,i}}{m}}, \\
 \mu_{jk} & = \frac{m_j m_k}{m_j+m_k} \;, \;\; 
 \mu_{jk,i} = \frac{m_i (m_j+m_k)}{m_i+m_j+m_k} \; ,
\end{align}
where $m$ is a normalization mass chosen to be the nucleon mass of
$939$~MeV$/c^2$.  The six hyperspherical relative coordinates are the
two pairs of directional angles, $(\Omega_{xi} , \Omega_{yi})$, for
$\mathbf{x}_i$ and $\mathbf{y}_i$, and hyperangle, $\alpha_i$, and 
hyperradial, $\rho$, coordinates defined by
\begin{align} \label{coord}
x_i = \rho \sin \alpha_i, && y_i = \rho \cos \alpha_i \;.
\end{align}

The Hamiltonian, $H$, can be written both in Jacobi and hyperspheric
relative coordinates.  For later convenience we focus here on the
Jacobi set where $\mathbf{x}_{cp}$ connects core and proton, that is
we can define
\begin{align}
 H &= H_x + H_y + V_{pp} \; ,  \label{eq62} \\   \label{eq64}
 H_x &= -\frac{\hbar^2}{2\mu_{cp}} \vec{\nabla}^2_{x_{cp}} + V_{cp} \; ,\\
  \label{eq66}
 H_y &=  - \frac{\hbar^2}{2\mu_{cp,p}} \vec{\nabla}^2_{y_{cp,p}} + V_{cp^{\prime}}  \; ,
\end{align}
where $\mu_{cp}$ and $\mu_{cp,p}$ are the reduced masses of
core-proton and proton to core-proton systems, respectively. The three two-body interactions are the proton-proton interaction, $V_{pp}$, the core-proton interaction, $V_{cp}$, and the interaction between the core and the second proton, $V_{cp^{\prime}}$. The coordinate dependencies of the two-body interactions,
$V_{cp}(\mathbf{x}_{cp})$,
$V_{cp^{\prime}}(\mathbf{x}_{cp},\mathbf{y}_{cp})$, and
$V_{pp}(\mathbf{x}_{cp},\mathbf{y}_{cp})$, between protons and core
are given as arguments.  In hyperspheric coordinates we have
\begin{align}
 T &= -\frac{\hbar^2}{2\mu_{cp}} \vec{\nabla}^2_{x_{cp}}
 - \frac{\hbar^2}{2\mu_{cp,p}} \vec{\nabla}^2_{y_{cp,p}}  =
 T_{\rho} + \frac{\hbar^2}{2 m \rho^2} \Lambda^2, \\
T_{\rho} &= - \frac{\hbar^2}{2 m} \left( \frac{\partial^2}{\partial \rho^2} + \frac{5}{\rho} \frac{\partial}{\partial \rho} \right)
  =  - \frac{\hbar^2}{2 m} \left( \rho^{-5}\frac{\partial}{\partial \rho}
  \rho^{5}  \frac{\partial}{\partial \rho}\right) \;, \\
\Lambda^2 &= - \frac{\partial^2}{\partial \alpha_i^2} -4 \cot (2\alpha_i) \frac{\partial}{\partial \alpha_i} + \frac{\hat{l}^2_{xi}}{\sin^2 \alpha_i} + \frac{\hat{l}^2_{yi} }{\cos^2 \alpha_i},
\end{align}
where $\hat{l}_{xi}$ and $\hat{l}_{yi}$ are the angular momentum
operators related to $\mathbf{x}_i$ and $\mathbf{y}_i$.

The method consists of an adiabatic expansion of the total
wave function $\Psi$, that is
\begin{align}
\Psi = \rho^{-5/2} \displaystyle\sum_n f_n(\rho) \Phi_n(\rho,\Omega)\;, \label{eq ang wave}
\end{align}
where each of the angular wave functions, $\Phi_n$, is a sum of Faddeev
components, $\Phi_n = \phi_{n,1} + \phi_{n,2} + \phi_{n,3}$, related
to the three corresponding Jacobi sets, and obeying the Faddeev equations
\begin{align}
0 &= \left( \Lambda^2 - \lambda_n \right) \phi_{n,i} +  \frac{2m}{\hbar^2} \rho^2 V_i \Phi_{n}. 
 \label{eq lamb}
\end{align}
The angular eigenvalues, $\lambda_n(\rho)$, and the related complete
set of eigenfunctions, $\Phi_{n}$, are first computed for each $\rho$.
Subsequently the radial expansion coefficients, $f_n(\rho)$, are found
from the coupled set of radial equations
\begin{align}
\bigg(-\frac{\partial^2}{\partial \rho^2} - \frac{2m E}{\hbar^2} &+ \frac{\lambda_n(\rho) + 15/4}{\rho^2} - Q_{nn} \bigg) f_n(\rho) \notag \\  
&= \displaystyle\sum_{n^{\prime}\neq n} \left( 2 P_{nn^{\prime}} \frac{\partial}{\partial \rho} + Q_{nn^{\prime}} \right) f_{n^{\prime}}(\rho), \\
P_{nn^{\prime}} &= \int \Phi^{\dagger}_n(\rho, \Omega) \frac{\partial}{\partial \rho} \Phi_{n^{\prime}} (\rho,\Omega) \, d\Omega,  \\
Q_{nn^{\prime}} &= \int \Phi^{\dagger}_n(\rho, \Omega) \frac{\partial^2}{\partial	\rho^2} \Phi_{n^{\prime}} (\rho,\Omega) \, d\Omega.
\end{align}
The left hand side reveals the crucial effective adiabatic part of the diagonal
potential acting on the particles
\begin{align}
 V_{eff,n} = \frac{\hbar^2}{2m} \left( \frac{\lambda_n(\rho) + 15/4}{\rho^2} 
 \right). \label{eq eff pot}
\end{align}

\subsection{Potentials and properties \label{sec pot}}

The result of the three-body calculations is dictated by the two-body
potentials employed.  The proton-proton interaction in free space is
well known in many details and with high accuracy. However, its influence on
the three-body solutions is only marginal provided, first of all, that the
$s$-wave scattering length is reproduced, and second that the
low-energy properties of the $p$ and $d$ partial waves are of
reasonable (small) size and in fair agreement with the
experimental values. The phenomenologically adjusted potentials described in Ref.~\cite{gar04} were used for the proton-proton interaction.

The test case used throughout this paper is the three-body system $^{68}\text{Se} + p + p$ ($^{70}\text{Kr}$). However, the results apply to the region in general, as the small mass and charge variations between the three critical waiting points are inconsequential for our purposes.

The proton-core potential is on the other hand decisive and a careful
choice has to be made.  For light-to-medium heavy cores of
mass numbers around $68$ we use the Woods-Saxon form with a spin-orbit
interaction, that is
\begin{align}
V(r) 
=& V_C(r) +   \frac{V_0}{1 + e^{(r -R)/a}} \notag \\ 
&+ \mathbf{l} \cdot \mathbf{s} \frac{1}{r} \frac{d}{d r}  
\frac{V_0^{ls}}{1 + e^{(r -R_{ls})/a_{ls}}},
\end{align}
where $r$ is the distance between the two bodies, $V_C$ is the Coulomb
potential, $V_0$ and $V_0^{ls}$ are the potential strengths of the
nuclear and spin-orbit potential, $\mathbf{l}$ and $\mathbf{s}$ are
the angular momentum and spin operators, and $R$ and $a$ are
parameters governing the radius and the thickness of the potentials.
This form is used for all partial waves, although the two strength
parameters vary strongly depending mainly on the angular momentum.  

A three-body potential will generally not be included. This would eventually be needed to adjust the energy levels according to a measured two- and three-body energy spectrum. However, the effect of such an addition is considered throughout the paper.

For the mass region in question we choose the values of the radial
shape parameter to be $R = 7.2$~fm, $R_{ls} = 6.3$~fm, $a=0.65$~fm,
and $a_{ls} = 0.5$~fm.  These choices are motivated by the knowledge
of the average nuclear mean-field potentials and densities
\cite{sie87}.  Accurate values are not needed because adjustments of
the strengths in any case are necessary for fine-tuning of the
two-body energies.  This is also one reason for omitting more
complicated spin-dependence like the tensor or quadratic spin-orbit
potentials.  In addition, the Coulomb potential is chosen from homogeneous charge distributions of radii, $5.6$~fm, and $1.8$~fm for core-proton system and proton-proton systems, respectively.

The two-body strength parameters are usually adjusted to reproduce the
core-proton bound and resonance energies, but unfortunately the energy spectrum of $^{69}$Br is not known experimentally. Based on shell model calculations the region is known to be near the midpoint of the $fpg$-shell. The most likely two-body orbitals are then $f_{5/2}$, $p_{3/2}$ (or possibly $p_{1/2}$), and $g_{9/2}$. However, recent experiment show that the $g_{9/2}$ orbital is not important for nuclei around $A=70$ with $N \simeq Z$ \cite{nic14}. Instead $f$ and $p$ orbitals are assumed to dominate the low-lying spectrum. This is also confirmed by the known spectrum of the mirror nuclei \cite{nes14}. To get opposite parity single-particle orbitals, necessary to form negative parity three-body states, a $d_{5/2}$ orbital can be included instead. To allow
occupancy only of these selected two-body core-proton states in the
three-body calculation is tricky because both lower- and higher-lying levels
must be excluded.  The Pauli forbidden two-body bound states are
excluded for each partial wave by use of shallow potentials without
bound states.  The large-distance properties are then precisely
correct but the unimportant nodes at small distances are then not
present for these excited states.

To locate one and only one partial wave at a given small energy we
must provide an accurately adjusted attractive potential, while all
other partial waves must have sufficiently strongly repulsive
potentials to prevent occupancy.  A given two-body energy then
provides one constraint correlating the two strengths.  To select one
and only one of two spin-orbit partners we choose a relatively high,
positive or negative, value of $V_0^{ls}$.  This may result in an
abnormal order of spin-orbit partners, but the goal is achieved.  The
different partial waves are completely independent of each other on
the two-body level, and we can therefore place all of them as we
choose.

\subsection{Radiative capture rate \label{sec radia}}

We want to calculate the waiting time before two protons from the
astrophysical environment are captured by a proton-dripline nuclear bound core.  The reaction rate, $R$, for the corresponding one-step $\gamma$, three-body transition
process, $p+p+c \rightarrow A + \gamma$, is given in general in Ref.~\cite{die11}. For the special system of two protons plus an even-even
core the rate becomes
\begin{align}
 R_{ppc}(E) = &  \frac{8 \pi}{(\mu_{cp} \mu_{cp,p})^{3/2}}  \frac{\hbar^3}{c^2} 
 \left( \frac{E_{\gamma}}{E} \right)^2 \sigma_{\gamma}(E_{\gamma}), 
\label{eq rate}
\end{align}
where $E_{\gamma} = E + B$ is the photon energy, $E$ the total three-body energy,
and $B$ is the three-body (positive) binding energy of the even-even
nucleus, $A$, with the wave function, $\Psi_{0}$.  The
photodissociation cross section, $\sigma_{\gamma}(E_{\gamma})$, for the
inverse process $A + \gamma \rightarrow a + b + c$, is a sum over
contributing electric multipole transitions of different orders,
$\ell$. That is
\begin{align}
\sigma^{\ell}_{\gamma}(E_{\gamma}) 
=& \frac{(2 \pi)^3 (\ell +1)}{\ell ((2\ell +1)!!)^2} \left( \frac{E_{\gamma}}{\hbar c}\right)^{2\ell-1} \frac{d}{d E}\mathcal{B}(\mathcal{E}\ell, 0 \rightarrow \ell), \label{eq siggam}
\end{align}
where the strength function for the $\mathcal{E}\ell$ transition,  
\begin{align}
 \frac{d}{d E} \mathcal{B}(\mathcal{E}\ell, 0 \rightarrow \ell) =
 \sum_i \left| \braket{\psi_{\ell}^{(i)} | | \hat{\Theta}_{\ell} | | 
 \Psi_{0}} \right|^2  \delta(E-E_i), \label{eq tran}
\end{align}
is given through the reduced matrix elements,
$\braket{\psi_{\ell}^{(i)} | | \hat{\Theta}_{\ell} | | \Psi_{0}}$,
where $\hat{\Theta}_{\ell}$ is the electric multipole operator,
$\psi_{\ell}^{(i)}$ is the wave function of energy, $E_i$, for all
bound and (discretized) three-body continuum states in the summation.

The astrophysical processes most often occur in a gas of given
temperature, $T$, which means we have to average the rate in
Eq.~(\ref{eq rate}) over the corresponding Maxwell-Boltzmann
distribution, $B(E,T) = \frac{1}{2} E^2 \exp(-E/T)/T^3$,
\begin{align}
\braket{R_{ppc}(E)} = \frac{1}{2T^3} \int E^2 R_{ppc}(E) \exp(-E/T)
 \, dE, \label{eq ave rate}
\end{align}
where the temperature is in units of energy. Combining Eqs.~(\ref{eq rate}), (\ref{eq siggam}), and (\ref{eq tran}) with Eq.~(\ref{eq ave rate}) results in a full three-body calculation of the energy-averaged reaction rate, as the wave functions in Eq.~(\ref{eq tran}) are proper three-body wave functions.

In the special case, where the resonances are very narrow and well separated the expression can be simplified greatly. If $\psi_{\ell}^{(i)}$ is a bound state, or a very narrow resonance state described accurately as a bound state, we can assign a photon emission width, $\Gamma_{\gamma}$, to the transition from this state, that is given by \cite{sie87}
\begin{align}
 \Gamma_{\gamma} = \frac{1}{2J+1} \frac{8 \pi (\ell +1)}{\ell ((2\ell+1)!!)^2}  
 \left(\frac{E_{\gamma}}{\hbar c}\right)^{2\ell+1}
\left| \braket{\Psi_{0}| | \hat{\Theta}_{\ell} | | \psi_{\ell}^{(i)} }
 \right|^2  \;,
 \label{eq gamgam}
\end{align}
where $J$ is the angular momentum of the three-body resonance with energy $E_R$.

If furthermore the three-body resonance is approximated by a Breit-Wigner shape, the photodissociation cross section in Eq.~(\ref{eq siggam}) can be written as
\begin{align}
\sigma_{\gamma}(E_{\gamma}) 
=& \pi (J+1/2)  \left( \frac{\hbar c}{E_{\gamma}} \right)^2  \frac{\Gamma_{eff}(E) \Gamma(E)}{(E-E_R)^2 + \frac{1}{4} \Gamma(E)^2}, \label{eq sig width}
\end{align}
where $(\Gamma_{eff}(E))^{-1} = (\Gamma_{ppc}(E))^{-1} + (\Gamma_{\gamma}(E))^{-1}$, $\Gamma_{ppc}$ is the strong decay
width, and the total (in principle energy dependent) width $\Gamma(E) = \Gamma_{ppc} + \Gamma_{\gamma}$.

With the dominating contributions arising from well separated, narrow resonances of Breit-Wigner shapes as in Eq.~(\ref{eq gamgam}), the integral in Eq.~(\ref{eq ave rate}) can be solved analytically, and we arrive at a very simple expression for the average rate 
\begin{align}
\braket{R_{ppc}(E)} =
 \frac{4 \pi^3 (2\ell+1)\hbar^5}{(\mu_{cp} \mu_{cp,p})^{3/2}}  
 \frac{\Gamma_{eff}(E_R)}{T^3} \exp(-E_R/T). \label{eq rate esti}
\end{align}
The only assumption in this expression is that the photodissociation
cross section is accurately expressed by the Breit-Wigner form in
Eq.~(\ref{eq gamgam}) with three-body resonance energy, $E_R \gg T$
and width $\Gamma \ll T$.  Other contributing narrow well-separated
resonances can simply be added.  Thus Eq.~(\ref{eq rate esti}) is
valid for one or more contributing narrow resonance irrespective of
capture mechanism.

The three-body (two proton plus an even-even core) calculation leads
to a $0^+$ ground state independent of the number and character of the
contributing core-proton single-particle states. The lowest excited three-body bound or resonance state would almost definitely be a $2^+$ state. This is confirmed by the mirror nuclei \cite{tul04}, where the two lowest excited states are $2^+$ states. Furthermore, the lowest possibly negative parity state is $2.5$ MeV above the ground state, and there is no $1^-$ at all in the known spectrum. The dominating transition can therefore be assumed to be an $\mathcal{E}2$ transition. To examine the remote possibility of a $1^-$ state two opposite parity single-particle state must be allowed. The relative energies of the $2^+$ and $1^-$ states depend on the attractions of the two-body potentials for the
corresponding partial waves.  The energies of these excited states
are, through Eq.~(\ref{eq rate esti}), all-decisive for the capture
rates.  Beside the three-body resonance energy also the effective
width, $\Gamma_{eff}$, is important for the capture rate.  The
realistic assumption that $\Gamma_{eff} \approx \Gamma_{\gamma}$
allows an estimate of the relative sizes of the $2^+$ to $1^-$
effective widths, that is from Ref.~\cite{sie87}
\begin{align}
\frac{\Gamma_{\gamma}(\mathcal{E}(\ell+1))}{\Gamma_{\gamma}(\mathcal{E}\ell)} 
\simeq \left( \frac{E_{\gamma} R }{\hbar c (2 \ell +3)} \right)^2, 
\label{eq wid rel}
\end{align}
where $R$ is the radius of the nucleus.  For $\ell = 1$ we get with
$E_{\gamma}=1$~MeV and $R=10$~fm that $\Gamma_{\gamma}(\mathcal{E}1)
\approx 10^{4} \, \Gamma_{\gamma}(\mathcal{E}2)$.  This estimate is from a
single-particle model, but within the three-body model effects from the other proton can at most contribute by a factor of two.  The implication is that $\mathcal{E}1$
would dominate unless forbidden by rather strict conservation
laws.

\section{Three-body ground state properties \label{sec 3b grnd}}

The nuclear properties of the waiting points are related to the borromean structure
reached after two-proton capture. We therefore first investigate the
borromean ground state depending on the two-body potentials varying
along the proton dripline.  In the following subsections we discuss
three-body energies and structure of the corresponding wave functions.

\subsection{Energies \label{sec engy}}

The most tightly bound borromean nucleus is most likely a closed (sub)shell for the
core while the additional two protons occupy empty valence orbits.  An
extra stability is present for $N=Z$ nuclei which conveniently also includes
waiting point nuclei for $N=Z= 32,34,36$.  The ground states for both
core and borromean nuclei have zero angular momentum and positive
parity as always for even-even nuclei.  The first excited states of
these nuclei are $0.9$~MeV \cite{sin07}, $0.9$~MeV \cite{mcc12} and
$0.7$~MeV \cite{abr10}, respectively.  These rather high values
demonstrate the stability required for a borromean three-body
structure build on these inert cores.  

Given the significant Coulomb barrier the three-body system does in principle not need to be strictly bound. The only requirement is that the proton-decay branch is small compared to the $\beta$-decay branch. Small positive three-body energies are then possible, which in principle extends our energy region of interest. However, the $\beta$ decay half-life of $^{70}$Kr has been measured to be 57(21) ms \cite{oin00}, so the possible extension is exceedingly small.

\begin{figure}[t]
\centering
  \includegraphics[width=1\linewidth]{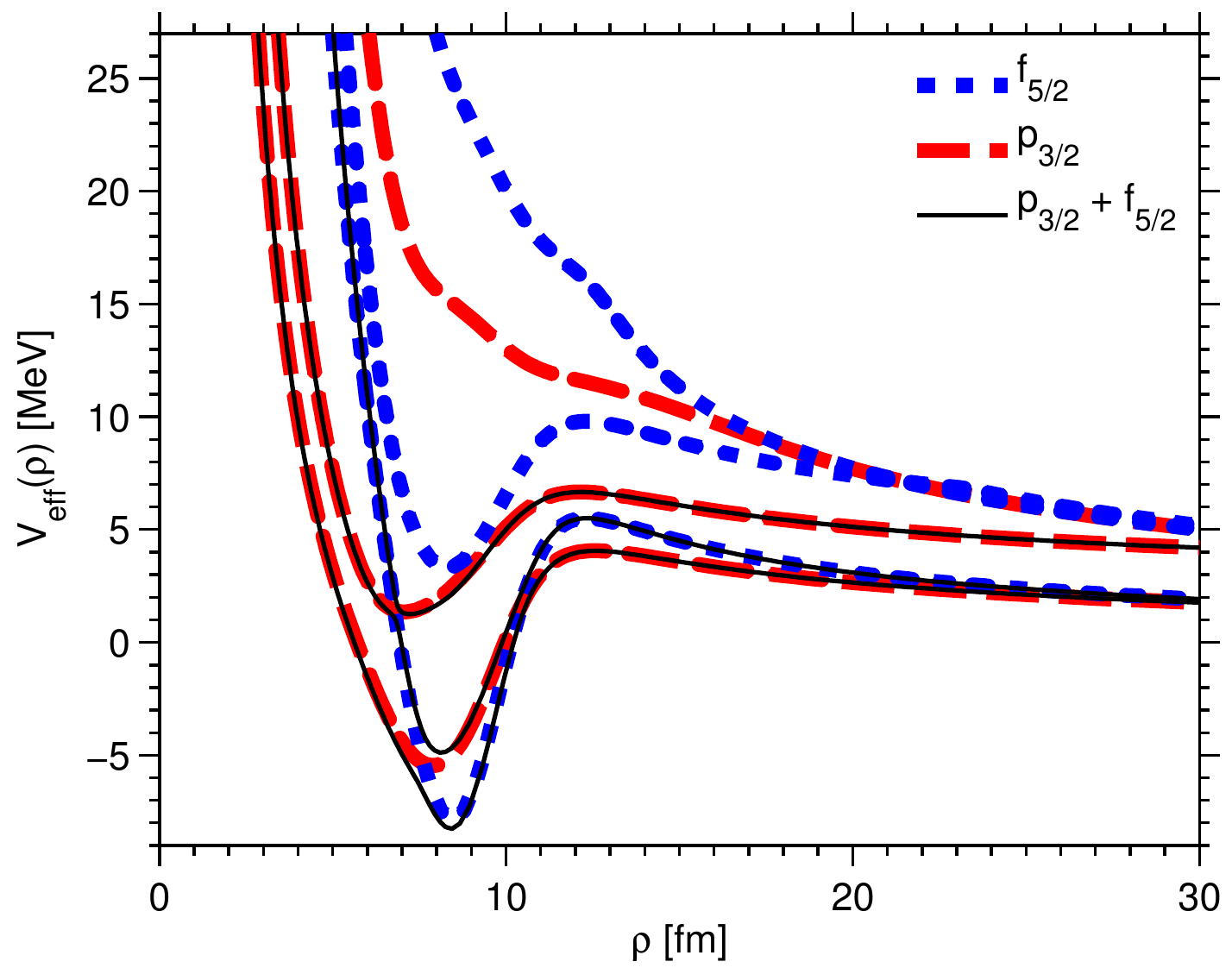}
  \caption{(Color online) The potentials based on the spectra of the lowest three $\lambda$'s for $p_{3/2}$ (dashed red line) and $f_{5/2}$ (dotted blue line) in isolation, along with the combination of both $p_{3/2}$ and $f_{5/2}$ (solid black line). All two-body potentials have been adjusted to produce the energy $E_{2b} = 0.1$ MeV. It is seen how the spectrum of the combined case always follows the lowest available potential. \label{fig pot comp}}
\end{figure}

The two-body bound or resonance energies for the different partial
waves are now the only pieces of information missing before the
three-body properties can be calculated.  We choose to use $p_{3/2}$
and $f_{5/2}$ and allow them both simultaneously with the same energy
as well as one at a time.  All other partial waves are either not
present or shifted to high energies. 

In Fig.~\ref{fig pot comp} we show the lowest effective adiabatic potentials defined in Eq.~(\ref{eq
  eff pot}) for the three cases with the same two-body energy, $E_{2b}
= 0.1$ MeV, in all cases. We first notice an attractive potential for each contributing
partial wave.  When both $p_{3/2}$ and $f_{5/2}$ are allowed we find
two attractive low-lying adiabatic potentials.  Both contributions
are present in the combined case where the deepest potential always
follow the lowest potential from either $p_{3/2}$ or $f_{5/2}$ for
different $\rho$-values.  Constructive (or destructive) interference
can only marginally occur through non-adiabatic terms since $p$ and
$f$-waves cannot couple to form a $0^+$ state.  However, this results in
avoided crossings for the combined case which therefore must produce
lower three-body energies than the individual cases.  In general, the
combined case can then never be less attractive than any of its
components.

\begin{figure}[t]
\centering
  \includegraphics[width=1\linewidth]{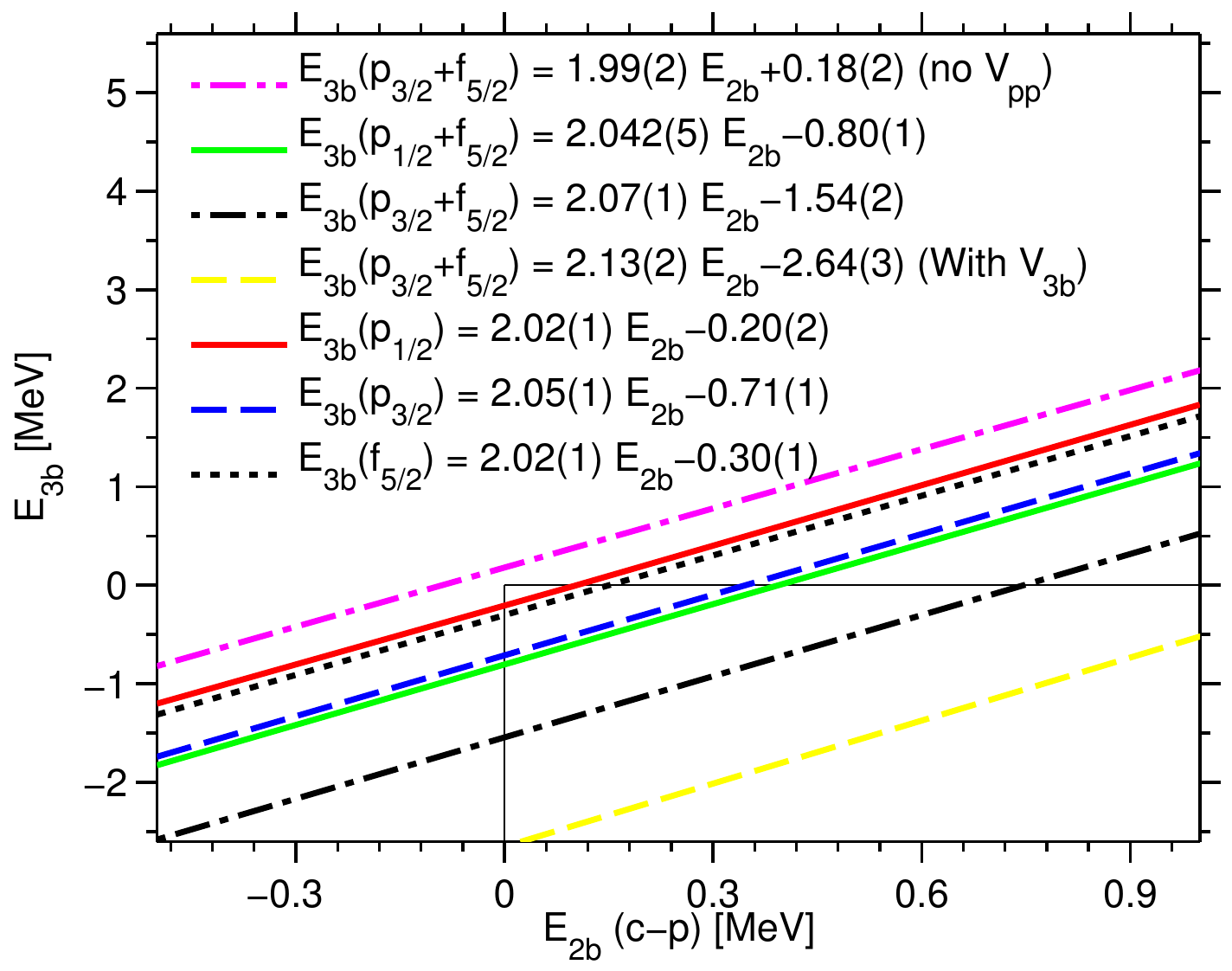}
  \caption{(Color online) The three-body energies as a function of two-body energies for $p_{1/2}$, $p_{3/2}$, and $f_{5/2}$ in isolation, and for $p_{1/2}$ and $f_{5/2}$ as well as $p_{3/2}$ and $f_{5/2}$ in combination, the latter both with and without p-p interaction, and with and without a three-body potential. The linear fits are in accordance with Eq.~(\ref{eq e2e3}). All two-body potentials were adjusted to produce the same energy, when more that one wave is allowed. Only a $0^+$ state is considered here. The horizontal and vertical line indicates our borromean region of interest. \label{fig ener}}
\end{figure}

The coupled set of adiabatic radial potentials is solved and energies
and wave functions calculated.  The two-body input parameters are not
known although limits from the required borromean character can be found.
We must therefore investigate the three-body properties as function of
the unknown two-body energies.  The three-body energies,
$E_{3b}$, are presented in Fig.~\ref{fig ener} as functions of two-body energies,
$E_{2b}$, for various selection of contributing partial waves.  The
most spectacular observation is that all curves are linear.  In the
figure we only exhibit results for the most interesting energy
interval but the same observed linear dependence is accurately
followed within the investigated interval,
$-2.0$~MeV~$<E_{2b}<2.0$~MeV.

When more than one partial wave is allowed, the two-body potentials
are adjusted to produce the same energy.  Fixing the energy of one two-body partial wave, while
increasing another one from the same value, the $E_{3b}$ must increase
and approach the higher-lying curve corresponding to the one contributing wave.
If more than two degenerate core-proton single-particle levels
contribute we would find even lower-lying $E_{3b}$-curves. However,
this is highly unlikely since this requires nuclear potentials of
unprecedented high symmetry.  Omission of the proton-proton
interaction in the case of two contributing partial waves, $p_{3/2}$
and $f_{5/2}$, increase the curve even above $(0,0)$, which apart from
center-of-mass effects should correspond to three non-interacting
systems.  Thus, we claim to have established limits for the
$E_{3b}$-variation between the lowest curve in Fig.~\ref{fig ener} and
a parallel curve roughly passing through $(0,0)$. 

So far, the results shown in Fig.~\ref{fig ener} have been obtained using just the two-body interactions contained in Eqs.~(\ref{eq62}), (\ref{eq64}), and (\ref{eq66}). However, it is a well known fact that using only two-body potentials will usually lead to an underbound three-body structure, as shown for instance for $^{17}$Ne and $^{12}$C in \cite{gar04b} and \cite{gar15b}, respectively. This problem is typically overtaken by addition of an effective three-body force, $V_{3b}(\rho)$, to the adiabatic potential given in Eq.~(\ref{eq eff pot}). Nevertheless, inclusion of such three-body force does not change the linear dependence shown in the figure. This is illustrated by the dashed (yellow) line in Fig.~\ref{fig ener}, which corresponds to the dot-dashed (black) case, but where a modest gaussian three-body force $V_{3b}(\rho)=S_{3b} \exp(-(\rho/\rho_0)^2)$ has been included. In particular, the values $S_{3b}=-5.8$ MeV and $\rho_0=6$ fm have been used. As seen in the figure, inclusion of the three-body force only shifts the three-body energy, but keeps the energy relation intact.

\subsection{Structure \label{sec struc}}
 
The simple linear dependency in Fig.~\ref{fig ener} is due to the
special structure of the wave function and the Hamiltonian in
Eq.~(\ref{eq62}).  The core is much heavier than the proton, and the
reduced masses are to a very good approximation both equal to the
proton mass.  Then ${x}_{cp} \approx {y}_{cp}$ and the Hamiltonians in
Eqs.~(\ref{eq64}) and (\ref{eq66}) are also approximately equal, $H_x
\simeq H_y$.  If $\ket{\Psi_{3b}}$ denotes the three-body
wave function, the three-body energy is determined as the expectation
value of the Hamiltonian in Eq.~(\ref{eq62}), that is
\begin{align}
E_{3b} 
&= \bra{\Psi_{3b}} H_x \ket{\Psi_{3b}} + \bra{\Psi_{3b}} H_y \ket{\Psi_{3b}} 
+ \bra{\Psi_{3b}} V_{pp} \ket{\Psi_{3b}} \notag \\
&\simeq 2E_x + E_{pp}.
\end{align}
where the approximate equality arises due to the assumption ${x}_{cp}
\approx {y}_{cp}$.  This approximation is consistent with
$\ket{\Psi_{3b}}$ as a product of corresponding two-body
wave functions.  For instance, when two single-particle states are
allowed, for example $p_{3/2}$ and $f_{5/2}$, we have $\ket{\Psi_{3b}}
= A \ket{p_{3/2}}_x \ket{p_{3/2}}_y + B \ket{f_{5/2}}_x
\ket{f_{5/2}}_y$, where the angular momentum coupling to $0^+$ of both
terms implicitly is assumed.  Product terms of $p_{3/2}$ and $f_{5/2}$ are not
allowed because the total angular momentum, $0^+$, cannot be reached.

The above product structure of $\ket{\Psi_{3b}}$ is rather general,
since the $\mathbf{x}_{cp}$ distribution necessarily is described by
the allowed single-particle wave functions whereas the same
overwhelmingly dominating single-particle $\mathbf{y}_{cp}$-part for
completeness should be extended to include other angular momentum
components.  We then get
\begin{eqnarray}
 E_x = \braket{\Psi_{3b}| H_x |\Psi_{3b}} = A^2 E_{2b}(p) + B^2 E_{2b}(f). 
 \label{eq mat ele}
\end{eqnarray}
where the two-body energies are defined by
\begin{eqnarray}
E_{2b}(p) &\equiv& \braket{{p_{3/2}}| H_x | {p_{3/2}}}_x\; , \nonumber \\
E_{2b}(f) &\equiv& \braket{{f_{5/2}}| H_x | {f_{5/2}}}_x \; ,
\label{eq two en}
\end{eqnarray}
and the cross terms vanish both due to orthogonality of the eigenfunctions
of $H_x$, and angular momentum conservation of the Hamiltonian.  We then arrive at the estimate of the three-body energy
\begin{align}
 E_{3b} \simeq 2 \left( A^2 E_{2b}(p) + B^2 E_{2b}(f) \right) + 
 \braket{\Psi_{3b} | V_{pp} | \Psi_{3b}} \;,
\label{eq e2e3}
\end{align}
where $A^2 + B^2 = 1$.  The occupation probabilities, $A^2$ and $B^2$,
are given by the relative weights of the partial waves. If only
$p_{3/2}$ or $f_{5/2}$ are allowed we have $A=1$ and $B=0$ or $A=0$ and
$B=1$, and consequently the linear dependence seen in Fig.~\ref{fig
  ener}.  If $E_{2b}(p)=E_{2b}(f)$ the linear dependence still
results, and in both extremes the slopes of the curves are two.
Variation between limits must produce a curve
connecting the corresponding two lines. For example, the lower
limit, when allowing both $p_{3/2}$ and $f_{5/2}$, is given by the
black, dash-dotted line, and the upper limit is given by the curve
corresponding to the wave with lowest energy. We emphasize the remarkable
equality of the slopes of the lines in Fig.~\ref{fig ener}.

\begin{table}
\centering
\caption{The weights of each partial wave for the same cases as shown in Fig.~\ref{fig ener}. The state in question is a $0^+$ state. Here $l_x$ denotes the relative angular momentum between the two particles specified by the second column, $l_y$ denotes the angular momentum of the third particle relative to the center of mass of the first two particles, and $l_t$ is the total angular momentum they combine to. The sixth column gives the order, $K_{max}$, of the Jacobi polynomium used for the corresponding partial wave.  The last two columns shows the weight (in percent) of the states for different two-body c-p energy. Components where all states have a weight less than $10\%$ are generally not included. \label{tab 3b wave}}
\begin{ruledtabular}
\begin{tabular}{ *{8}{c}}
                         &          &      &      &      &           & \multicolumn{2}{c}{$E_{2b}$ (MeV)} \\
          \cline{7-8}
Waves                    &   Jacobi & $l_x$& $l_y$& $l_t$& $K_{max}$ & -2.0 &  2.0 \\
\colrule
$p_{3/2}$                & p-p      &  0   &  0   &  0   &   98  &   78  &   76 \\
                         &          &  1   &  1   &  1   &   80  &   20  &   22 \\
                  \cline{3-8}
                         &  p-c     &  1   &  1   &  0   &   80  &   79  &   77 \\
                         &          &  1   &  1   &  1   &   80  &   20  &   23 \\
                  \cline{2-8}
$p_{1/2}$                & p-p      &  0   &  0   &  0   &   98  &   48  &   45 \\
                         &          &  1   &  1   &  1   &   80  &   49  &   52 \\
                  \cline{3-8}
                         &  p-c     &  1   &  1   &  0   &   80  &   49  &   46 \\
                         &          &  1   &  1   &  1   &   80  &   51  &   54 \\
                  \cline{2-8}
$f_{5/2}$                & p-p      &  0   &  0   &  0   &   98  &   48  &   47 \\
                         &          &  1   &  1   &  1   &   80  &   37  &   38 \\
                         &          &  2   &  2   &  0   &   52  &   11  &   11 \\
                  \cline{3-8}
                         &  p-c     &  3   &  3   &  0   &   74  &   60  &   59 \\
                         &          &  3   &  3   &  1   &   74  &   40  &   41 \\ 
                  \cline{2-8}
$p_{3/2}+f_{5/2}$        & p-p      &  0   &  0   &  0   &   98  &   78  &   75 \\
                         &          &  1   &  1   &  1   &   80  &   19  &   22 \\
                  \cline{3-8}
                         &  p-c     &  1   &  1   &  0   &   80  &   56  &   52 \\
                         &          &  1   &  1   &  1   &   80  &   12  &   13 \\
                         &          &  3   &  3   &  0   &   74  &   23  &   24 \\
                         &          &  3   &  3   &  1   &   74  &    9  &   11 \\
                  \cline{2-8}
$p_{3/2}+f_{5/2}$        & p-p      &  0   &  0   &  0   &   98  &   80  &   77 \\
$(V_{3b} \neq 0)$        &          &  1   &  1   &  1   &   80  &   18  &   20 \\
                  \cline{3-8}
                         &  p-c     &  1   &  1   &  0   &   80  &   62  &   57 \\
                         &          &  1   &  1   &  1   &   80  &   13  &   13 \\
                         &          &  3   &  3   &  0   &   74  &   19  &   21 \\
                         &          &  3   &  3   &  1   &   74  &    6  &    9 \\                         
                  \cline{2-8}
$p_{3/2}+f_{5/2}$        & p-p      &  0   &  0   &  0   &   98  &   65  &   64 \\
$(V_{pp}=0)$             &          &  1   &  1   &  1   &   80  &   33  &   32 \\
                  \cline{3-8}
                         &  p-c     &  1   &  1   &  0   &   80  &   66  &   65 \\
                         &          &  1   &  1   &  1   &   80  &   34  &   33 \\
                         &          &  3   &  3   &  0   &   74  &    0  &    0 \\
                         &          &  3   &  3   &  1   &   74  &    0  &    0 \\
                  \cline{2-8}
$p_{1/2}+f_{5/2}$        & p-p      &  0   &  0   &  0   &   98  &   66  &   61 \\
                         &          &  1   &  1   &  1   &   80  &   31  &   35 \\
                  \cline{3-8}
                         &  p-c     &  1   &  1   &  0   &   80  &   38  &   34 \\
                         &          &  1   &  1   &  1   &   80  &   20  &   21 \\
                         &          &  3   &  3   &  0   &   74  &   30  &   31 \\
                         &          &  3   &  3   &  1   &   74  &   11  &   14 \\                         
\end{tabular}
\end{ruledtabular}
\end{table}

The structure of the solutions is quantified by decomposition into
contributing partial waves.  We collect the results in Table \ref{tab 3b wave} for the same seven cases as seen in Fig.~\ref{fig ener} with both sets of Jacobi
coordinates.  We have chosen two rather different values, $-2.0$~MeV
and $2.0$~MeV, for the two-body energies used to adjust the strength
parameters.  The distributions are very similar for the two energies,
and the transition from one distribution to the other is slow and
monotonous. Adding a three-body potential has no significant effect on the structure of the wave function.

The decomposition is rather trivial, when only one single-particle
orbit is allowed.  In the second set of Jacobi coordinates, the
proton-core set, only this state is allowed.  However, two couplings to
total orbital angular momentum, $l_t = 0$ and $1$, share the weights,
which is equally distributed in the first Jacobi set since $l_t$ is
conserved in this rotation.

When two single-particle orbits are allowed the decomposition includes
both structures with different relative weights. The $l_t=0$
components are always far larger than those of $l_t=1$.  The
distribution is strongly depending on the proton-proton interaction as
seen by the complete separation for $V_{pp}=0$, where the three-body
ground state is degenerate corresponding to either $\ket{p_{3/2}}_x
\ket{p_{3/2}}_y$ or $\ket{f_{5/2}}_x \ket{f_{5/2}}_y$.

As seen from Eq.~(\ref{eq e2e3}) the displacement of the lines in Fig.~\ref{fig ener} not only depends on the proton-proton interaction, but also on the wave function. The spin-orbit difference between $p_{1/2}$ and $p_{3/2}$ causes a slightly different wave function, which is why the two curves differ. This is also demonstrated by the fact that displacement changes from $-1.54$ to $0.18$~MeV, when a very shallow nuclear potential and neither spin-spin, spin-orbit, tensor, or Coulomb potentials are used in the proton-proton interaction.  In fact, by completely eliminating the proton-proton interaction, the result is almost two independent two-body systems. The slight positive displacement is in that case due to the fact that all three coordinates $\mathbf{r}_1$, $\mathbf{r}_2$, and $\mathbf{r}_3$ are still coupled in the $\mathbf{y}$ coordinate as long as the core is not infinitely heavy.

\begin{figure}[t]
\centering
  \includegraphics[width=1\linewidth]{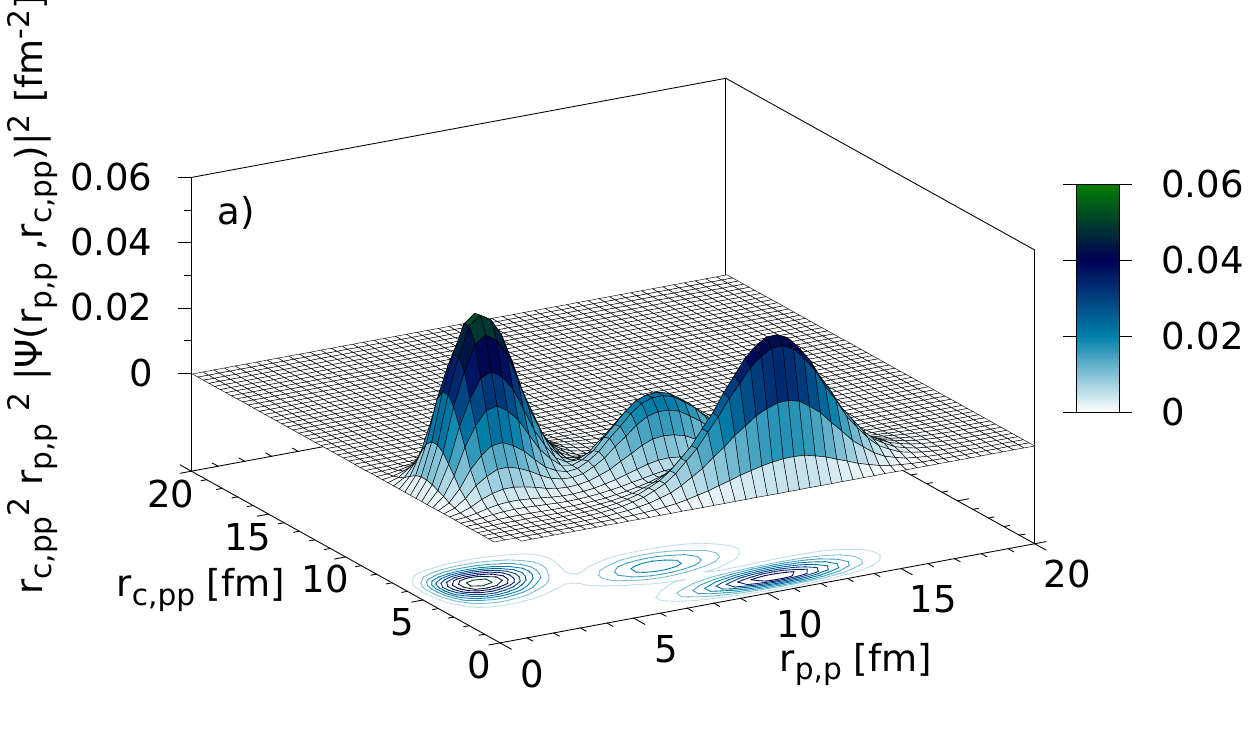}
  \includegraphics[width=1\linewidth]{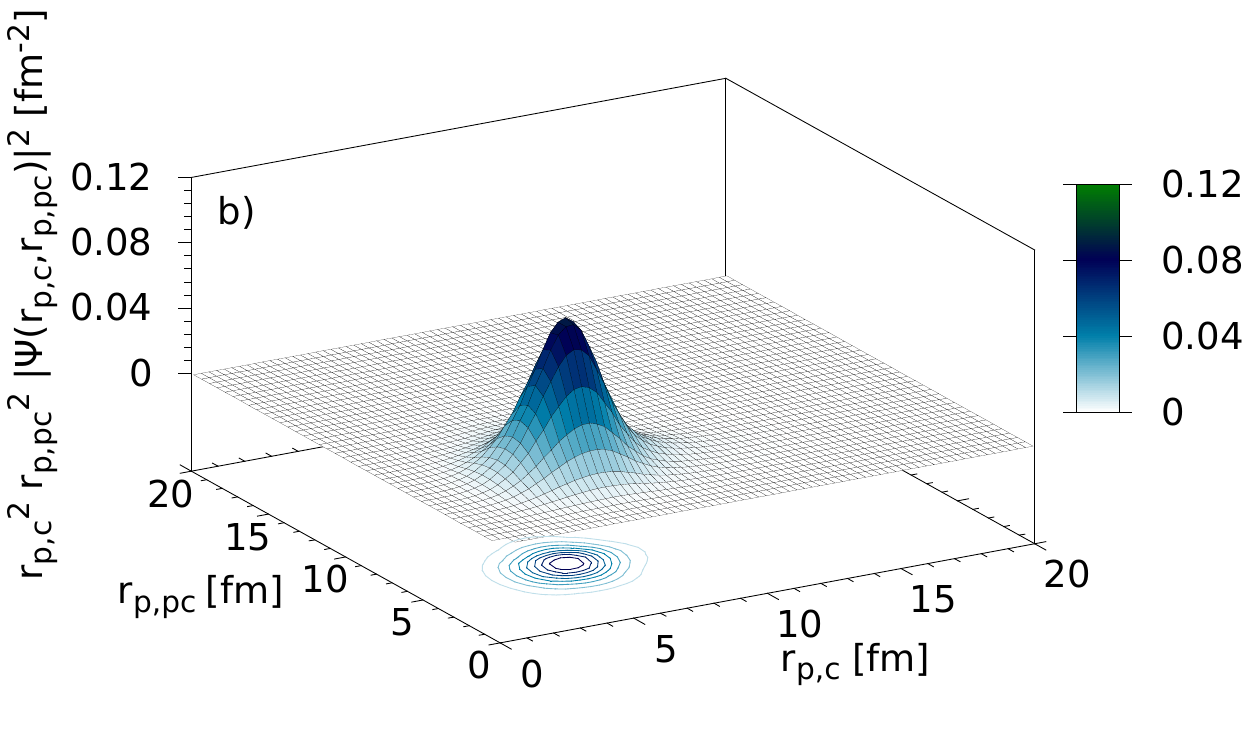}
  \caption{(Color online) The probability distribution for the $0^+$
    ground state with equal two-body binding of 0.641 MeV for both $f_{5/2}$ and
    $p_{3/2}$ partial waves.  Projected contour curves are shown at
    the bottom of each figure.  The distance variables correspond to
    the two different Jacobi sets, where the a) and b) panels are for
    the first and second Jacobi sets, respectively.
\label{fig ppdens}}
\end{figure}

The actual size of the displacement, caused by the proton-proton
interaction, can in principle be estimated from the matrix element
$\braket{\Psi | V_{pp} | \Psi}$.  We show the probability distribution of the three-body wave function
in Fig.~\ref{fig ppdens} where the distance of about $5$~fm between
proton and core (Fig.~\ref{fig ppdens}b) is a prominent feature. The proton-proton
distance distribution (Fig.~\ref{fig ppdens}a) is much more complicated with three peaks
at distances of about $2.5$~fm, $8$~fm, and $11$~fm, respectively.
The Coulomb repulsion would correspondingly be about $0.6$~MeV,
$0.18$~MeV, and $0.13$~MeV.  

The strong nucleon-nucleon interaction has strength of around $40$~MeV and range of
$2$~fm. To arrive at a total displacement of around $-1.7$~MeV there
must be only a few percent of the proton-proton distance-probability
within the range of $2$~fm.  To make a reliable estimate of the displacement is therefore very
delicate as it depends strongly on the solution to the three-body
problem. A much better computation would be to evaluate $E_{pp}$
directly.  However, this would still be only an estimate since $\Psi$
changes with the interaction, and at best we can only reproduce the
already known actual curves in Fig.~\ref{fig ener}.

\section{Radiative capture \label{sec pro}}

The critical waiting points in the rp process are
defined by the long time it takes to capture an additional two protons in the nucleus. The borromean nature requires a three-body reaction producing the strong interaction bound two-proton plus core system.
In this section we shall focus on corresponding reaction rates and the
structure of crucial intermediate states.  The total process is $c + p
+ p \rightarrow A + \gamma$, which as well can be understood through
the reverse process, $A + \gamma \rightarrow c + p + p$.  It is often
very accurate to divide part of this process into two steps, $A^*
\leftrightarrow A + \gamma$, where $A^*$ is one (and sometimes a few)
intermediate excited state.  We shall first investigate the
properties of such intermediate excited states and subsequently
calculate the reaction rates. Unless otherwise stated a two-body energy of 0.641 MeV is used, as this is the measured proton separation energy of $^{69}\text{Br}$ \cite{san14}. The results are not restricted to such a specific energy, but apply rather generally to the region of the nuclear chart around the critical waiting points.

\subsection{Excited continuum states \label{sec cont}}

The reactions proceed from continuum three-body states, that is from two
free protons and an (almost) ordinary nucleus. This problem can be
handled by two conceptually different methods \cite{gar15a,gar15b}, where the
first is to specify the boundary conditions directly and solve the
Schr\"{o}dinger equation.  The second method is to discretize the
continuum in a large hyperradial box which in the present case is
limited by hyperradii less than the box radius, $\rho_{max}$.

\begin{figure}[tb!]
\centering
  \includegraphics[width=1\linewidth]{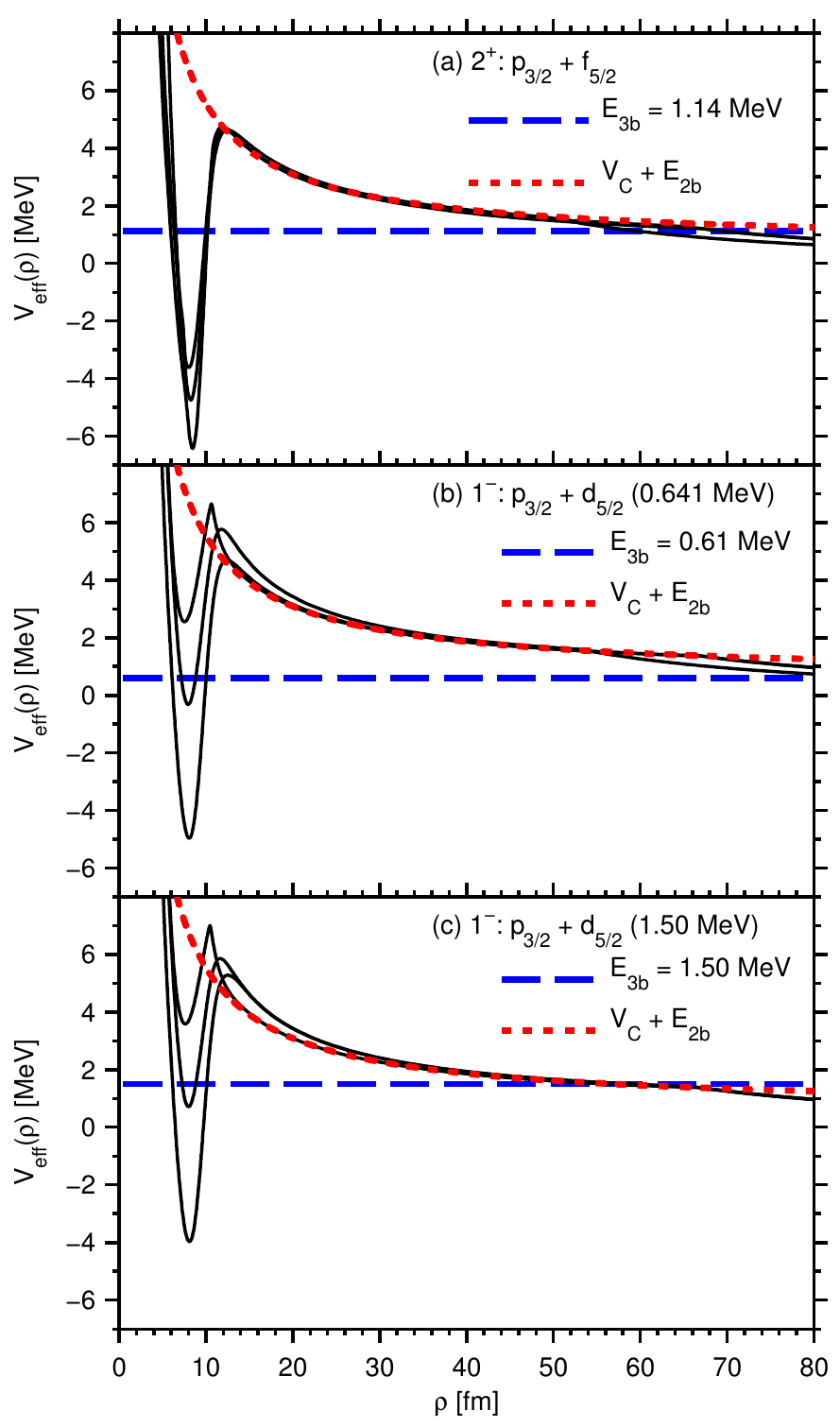}
  \caption{(Color online) (a) The potentials based on the spectra of
    the lowest five $\lambda$'s for the $2^+$ state allowing both
    $p_{3/2}$ and $f_{5/2}$ waves. The dashed, horizontal line is at
    the lowest three-body resonance energy. The dotted curve is the sum of
    Coulomb potential and two-body energy. (b) The same for the $1^-$
    state with $p_{3/2}$ and $d_{5/2}$ waves. (c) The
    same as for (b) with the $d_{5/2}$ two-body energy equal to $1.5$~MeV.
\label{fig pot 2+}}
\end{figure}

We shall here use the discretization method with the great advantage of using the already defined adiabatic potentials.  As discussed in Sec.~\ref{sec radia} the most likely dominating intermediate angular momenta is $2^+$. Also, as discussed in Sec.~\ref{sec pot}, states with the necessary opposite parity to form a $1^-$ state is also very unlikely. However, to later study the scale of the $\mathcal{E}1$ transition, $p_{3/2}$ and $d_{5/2}$ waves are allowed to form a $1^-$ state. We show the lowest corresponding adiabatic potentials in Fig.~\ref{fig pot 2+} for three such cases, where two partial waves for each total angular momentum are allowed. The parameters are chosen to be the same as already used in the study of ground state properties.

The potentials in (a) and (b) of Fig.~\ref{fig pot 2+} are
qualitatively similar to those of Fig.~\ref{fig pot comp} with
attractive pockets around $\rho = 8$~fm, infinite repulsion at smaller
distances, and barriers at larger $\rho$ separating regions of
interacting and fully separated three particles.  The attractive
region and the substantial barrier suggest that there are narrow, low-lying
resonances or perhaps even bound states.  The energies of the radial
solutions in the box are at first glance also similar to the ground
state $0^+$ solutions, that is one prominent bound, separate, low-lying
state and a number of higher-lying solutions.  However, the bound
solutions have positive energy and would therefore correspond to resonances
or continuum states. The $1^-$ potentials shown as (c) in Fig.~\ref{fig pot 2+} are much more repulsive at short distances since the necessary $d_{5/2}$ state is chosen to be at 1.50 MeV. The resonance is at a higher energy, where also more continuum background states have a non-vanishing contribution at short distances. 

The three-body energy for the $1^-$ case with $E_{2b}(p) \neq E_{2b}(d)$ can be explained using an argument similar to that in Sec.~\ref{sec struc}. The three-body wave function can be written as $\ket{\Psi_{3b}} = A \ket{p_{3/2}}_x \ket{d_{5/2}}_y + B \ket{d_{5/2}}_x \ket{p_{3/2}}_y$, where the angular momentum coupling is to $1^-$. As $m_c \gg m_p$ the likely weight of the two configurations must be the same, and since these are the only possibilities we have $A^2 = B^2 = 1/2$. Then $E_x = 1/2 (E_{2b}(p_{3/2}) + E_{2b}(d_{5/2}) )$, and $E_{3b} = E_{2b}(p_{3/2}) + E_{2b}(d_{5/2}) + E_{pp}$. It was found that if $E_{2b}(p_{3/2}) = E_{2b}(d_{5/2}) = 0.641$ MeV then $E_{3b} = 0.61$ MeV, which means $E_{pp} \sim -E_{2b}(p_{3/2})$ for that specific energy. However, the three-body wave function is almost unchanged when changing $E_{2b}(d_{5/2})$ because both the p- and d-wave is needed to form the $1^-$ state, and $E_{pp}$ is therefore also almost unchanged. The result is $E_{3b} \simeq E_{2b}(d_{5/2})$ for $E_{2b}(p_{3/2}) = 0.641$ MeV. 

The three-body large-distance configuration is also revealed by close inspection of Fig.~\ref{fig pot 2+}. A region around $\rho=40-50$~fm is seen where one potential tend to be more flat than dictated by a general Coulomb decreasing potential.  The size and slow decrease is precisely consistent with the Coulomb potential between one proton and a proton-core two-body system in a spatially small resonance at $0.641$~MeV. This is reflected in the good agreement between the lowest potential and the corresponding Coulomb potential plus two-body energy seen in Fig.~\ref{fig pot 2+}. In Fig.~\ref{fig pot 2+}(c) the same $V_C+E_{2b}$ barrier is indicated by the red, dotted curve. This implies that the close-lying proton is in a p-orbital.

\begin{figure}[t]
\centering
  \includegraphics[width=1\linewidth]{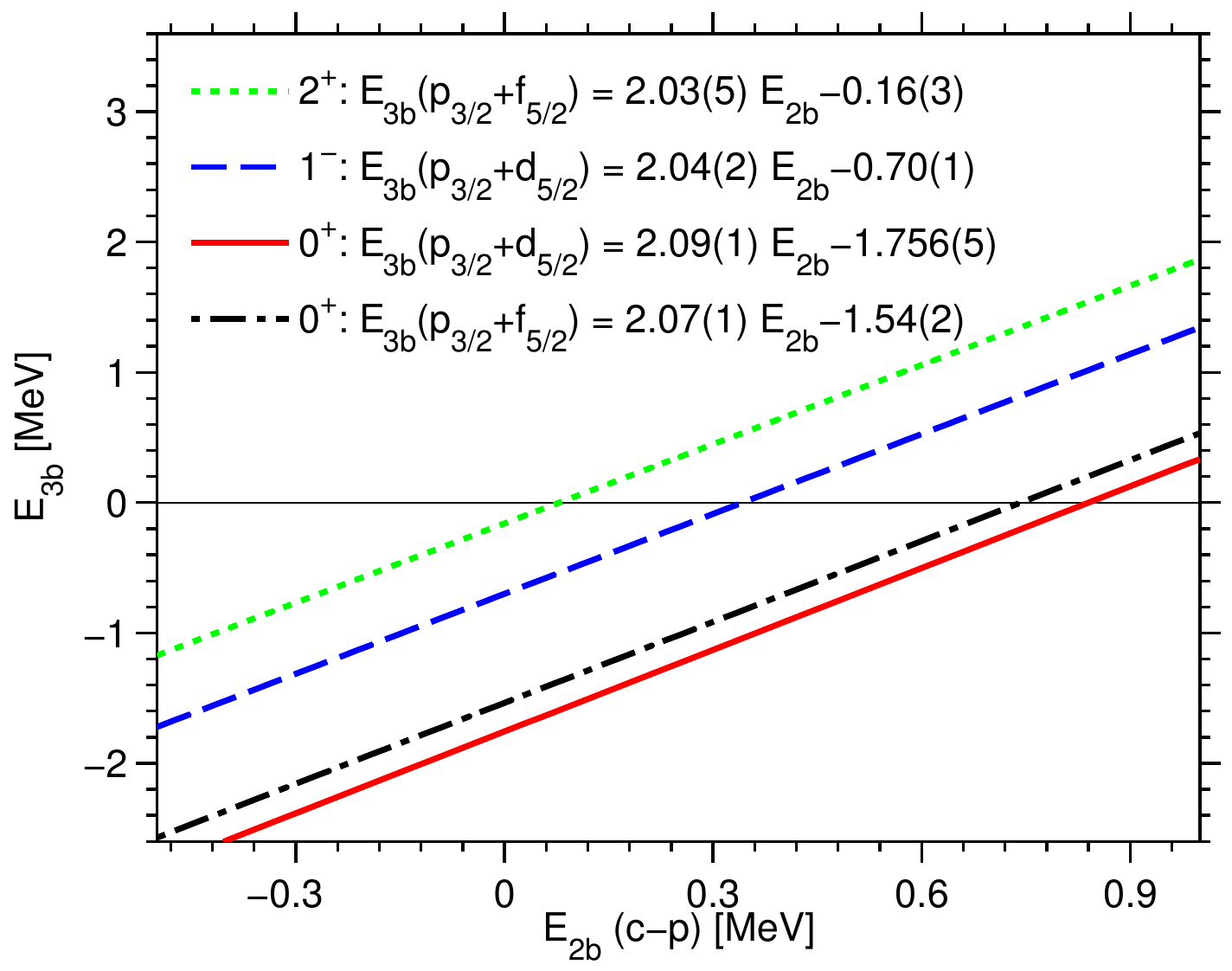}
  \caption{(Color online) The three-body energy as a function of two-body energy for the $0^+$ and $2^+$ state with $p_{3/2}$ and $f_{5/2}$ waves, as well as for the $0^+$ and $1^-$ state with $p_{3/2}$ and $d_{5/2}$ waves. The horizontal line is included to guide the eye.
\label{fig engy cont}}
\end{figure}

The three-body (bound) resonance energies are not necessarily the
lowest discretized states. It depends strongly on the size of the box
since an infinite box must produce a continuum of states from the
threshold and upwards.  We detect the resonance state by requiring that
the density distribution is localized at small distances. The resonance energies
are shown in Fig.~\ref{fig engy cont} as function of two-body energies
for the choices used for the ground state calculations.  It is perhaps
less surprising to find the same simple and accurate linear dependence
as we observed for the ground states.  The criteria for linearity are
the same. Also the effect of a three-body potential would be similar to the effect seen in Fig.~\ref{fig ener} only it would be more difficult to justify the exact size of the potential as the excitation spectrum is less well known.

\begin{table}[t]
\centering
\caption{The weights of each partial wave for the $2^+$ case, where only $p_{3/2}$ and $f_{5/2}$ waves are allowed, and for the $1^-$ case, where only $p_{3/2}$ and $d_{5/2}$ waves are allowed. The notation is the same as in Table \ref{tab 3b wave}. The last three columns shows the weight (in percent) of the lowest resonances. The resonances correspond to the first peaks in Fig.~\ref{fig cross}. Components where all states have a weight less than $10\%$ are not included. \label{tab 3b excwave}}
\begin{ruledtabular}
\begin{tabular}{l *{8}{c}}
Waves                       &   Jacobi & $l_x$& $l_y$& $l_t$& $K_{max}$ & \multicolumn{3}{c}{Weight}  \\
\colrule
$2^+$                       & p-p      &  0   &  2   &  2   &  199  &   27     &   8   &  15  \\
$p_{3/2}+f_{5/2}$           &          &  1   &  1   &  1   &  199  &   40     &  11   &  11  \\
                            &          &  1   &  1   &  2   &  199  &    3     &  19   &   2  \\
                            &          &  1   &  3   &  3   &  202  &    1     &  17   &  12  \\
                            &          &  2   &  0   &  2   &   60  &   23     &   8   &  13  \\
                            &          &  3   &  1   &  3   &   22  &    1     &  17   &  12  \\
                  \cline{3-9}
                            &  p-c     &  1   &  1   &  1   &  160  &   41     &  12   &  10  \\
                            &          &  1   &  1   &  2   &  160  &   38     &   4   &   1  \\
                            &          &  1   &  3   &  2   &  162  &    3     &  15   &  15  \\
                            &          &  1   &  3   &  3   &  162  &    1     &  20   &  23  \\
                            &          &  3   &  1   &  2   &  162  &    3     &  15   &  15  \\
                            &          &  3   &  1   &  3   &  162  &    1     &  20   &  23  \\
                  \cline{2-9}
$2^+$                       & p-p      &  0   &  2   &  2   &  199  &   23     &  25   &  --  \\
$p_{3/2}$                   &          &  1   &  1   &  1   &  199  &   53     &  41   &  --  \\
                            &          &  2   &  0   &  2   &   60  &   21     &  24   &  --  \\
                  \cline{3-9}
                            &  p-c     &  1   &  1   &  1   &  200  &   55     &  42   &  --  \\
                            &          &  1   &  1   &  2   &  200  &   45     &  52   &  --  \\
                  \cline{2-9}                  
$1^-$\footnote{$E_{2b}(p_{3/2}) = E_{2b}(d_{5/2}) = 0.641$ MeV.}
                            & p-p      &  0   &  1   &  1   &  199  &   66     &  --   &  --  \\
$p_{3/2}+d_{5/2}$           &          &  1   &  0   &  1   &  199  &   13     &  --   &  --  \\
                  \cline{3-9}
                            &  p-c     &  1   &  2   &  1   &  161  &   38     &  --   &  --  \\
                            &          &  2   &  1   &  1   &  161  &   38     &  --   &  --  \\
                  \cline{2-9}                  
$1^-$\footnote{$E_{2b}(p_{3/2}) = 0.641$ MeV, $E_{2b}(d_{5/2}) = 1.50$ MeV.}
                            & p-p      &  0   &  1   &  1   &  199  &   66     &  --   &  --  \\
$p_{3/2}+d_{5/2}$           &          &  1   &  0   &  1   &  199  &   14     &  --   &  --  \\
                  \cline{3-9}
                            &  p-c     &  1   &  2   &  1   &  161  &   38     &  --   &  --  \\
                            &          &  2   &  1   &  1   &  161  &   38     &  --   &  --  \\
\end{tabular}
\end{ruledtabular}
\end{table}

We can then conclude that the excitation energy is a constant
independent of the chosen variation in Figs.~\ref{fig engy cont} and
\ref{fig ener}.  Specifically we get excitation energies $1.38$~MeV and
$1.06$~MeV for the $2^+$ and the $1^-$ states,
respectively.  However, the derived limits of variation are completely
different.  The $2^+$ excitation energy is strongly limited, since the
ground state only can move between finite limits as shown in
Fig.~\ref{fig ener}, and the lowest $2^+$ resonance is entirely determined by
the $p_{3/2}$ component even when also $f_{5/2}$ is allowed with the
same energy as seen in Table~\ref{tab 3b excwave}.

The reason for this behaviour is that any partial wave (except angular momentum $1/2$)
of proton and $0^+$-core states can be occupied by two protons coupled
to both $0^+$ and $2^+$.  The excitation energy can then vary at most
by about $1.5$~MeV.  In contrast, the $1^-$ state requires both positive
and negative parity proton-core single-particle states.  Thus, a well
defined finite $0^+$ ground state energy always appears whereas the
$1^-$ excitation energy can vary from zero to infinity by increasing
one of the necessary proton-core states towards infinity.

The discretization allows normalization of all the states in the box.
Specifically, as mentioned above the density distributions of the
resonance states are localized at small distances.  The probability
distributions are almost indistinguishable from that of Fig.~\ref{fig
  ppdens} each with a peak at a proton-core distance of about $6$~fm.
Only one $1^-$ resonance is found whereas three rather pronounced low-lying
$2^+$ resonances appear when both $p_{3/2}$ and $f_{5/2}$ are allowed
with the same energies.  By definition all these states have the
overwhelmingly large probability located at small distances.  The
spatial overlaps with corresponding ground states are therefore very
large in all theses cases.  Thus, only appropriate angular momentum
dependent operators and resonance energies are required to initiate highly likely
transitions, as seen in Eq.~(\ref{eq siggam}).

It is then important to know the angular momentum composition of the
excited states.  The choices of allowed partial waves are strongly
limiting for these distributions as seen in Table~\ref{tab 3b
  excwave}.  The simplest are the lowest $2^+$ state and the $1^-$
resonance.  They consists of only proton-core $p_{3/2}$ components,
and equal measures of proton-core and $p_{3/2}$-$d_{5/2}$ components,
respectively.  The two excited $2^+$ states are mixtures of $p_{3/2}$
and $f_{5/2}$ proton-core partial wave components.

All discretized continuum states, beside the resonances discussed
above, are much more dilute and spread out at large distances of the
box.  The spatial overlaps with the ground state are therefore very
small and any transitions would correspondingly be reduced in size.
This does not necessarily mean that their contributions can be
ignored, because the number of these states also increase both with
box size and with energy.  At some point they overlap and contribute
as a genuine continuum.

\subsection{Cross section and reaction rates  \label{sec reac res}}

The three particles in the continuum are not characterized by one
complete set of quantum numbers. The plane wave states for free
particles contain all angular momenta in a partial wave expansion.  In
contrast the final nuclear state has given angular momentum and parity,
and the transition itself is conveniently specified by a given
one-body multipole operator.  The transitions between well defined
states are independent of each other, and the prescription is
therefore to calculate and add the different contributions.  The
transition probabilities decrease strongly with multipolarity, which
therefore is decided by nature through the structure of the borromean
final state.

The $0^+$ quantum numbers are achieved by coupling of the two
proton-core angular momenta, which must be unoccupied by core
nucleons.  The available low-lying single-particle orbits therefore
depends entirely on the region of interest in the nuclear chart.  We
focuss on the proton dripline region around $A=68$, and as discussed $f_{5/2}$ and $p_{3/2}$ are from the mirror nuclei expected to be the dominating single-particle orbitals, with $\mathcal{E}2$ being the dominating transition. For completeness we shall nevertheless investigate the heavily suppressed $\mathcal{E}1$ transition.

We proceed by calculating the discretized three-body continuum states
for given total angular momentum and parity with selected two-body
input energies.  The cross sections and reaction rates can then be
obtained by summation over these discrete ``continuum'' states.  The
method does not assume any specific reaction mechanism, that is two-step via a photon emitting resonance and/or continuum state, and both sequential and direct
reactions are included in the numerically obtained results.

The discretization implies that the low-energy spectrum can be too
sparse when the box size is comparable to the extension of the
potential barriers.  The space outside the barriers is then too small
to provide box bound states. On the other hand, an attractive
short-distance pocket would produce an isolated bound-state like
resonance which then would mediate all the low-energy transition
probability.  The missing continuum states in its energy neighbourhood
would have vanishing spatial overlap with the ground state, and
consequently also vanishing transition probability.  In the
high-energy limit the level density increases and the contributions
are distributed over many levels.  At some point the included Hilbert
space in the basis becomes insufficient.  Fortunately, we are able to
cover an energy region sufficient for the astrophysical reactions of
interest.

\begin{figure}[t]
\centering
  \includegraphics[width=1\linewidth]{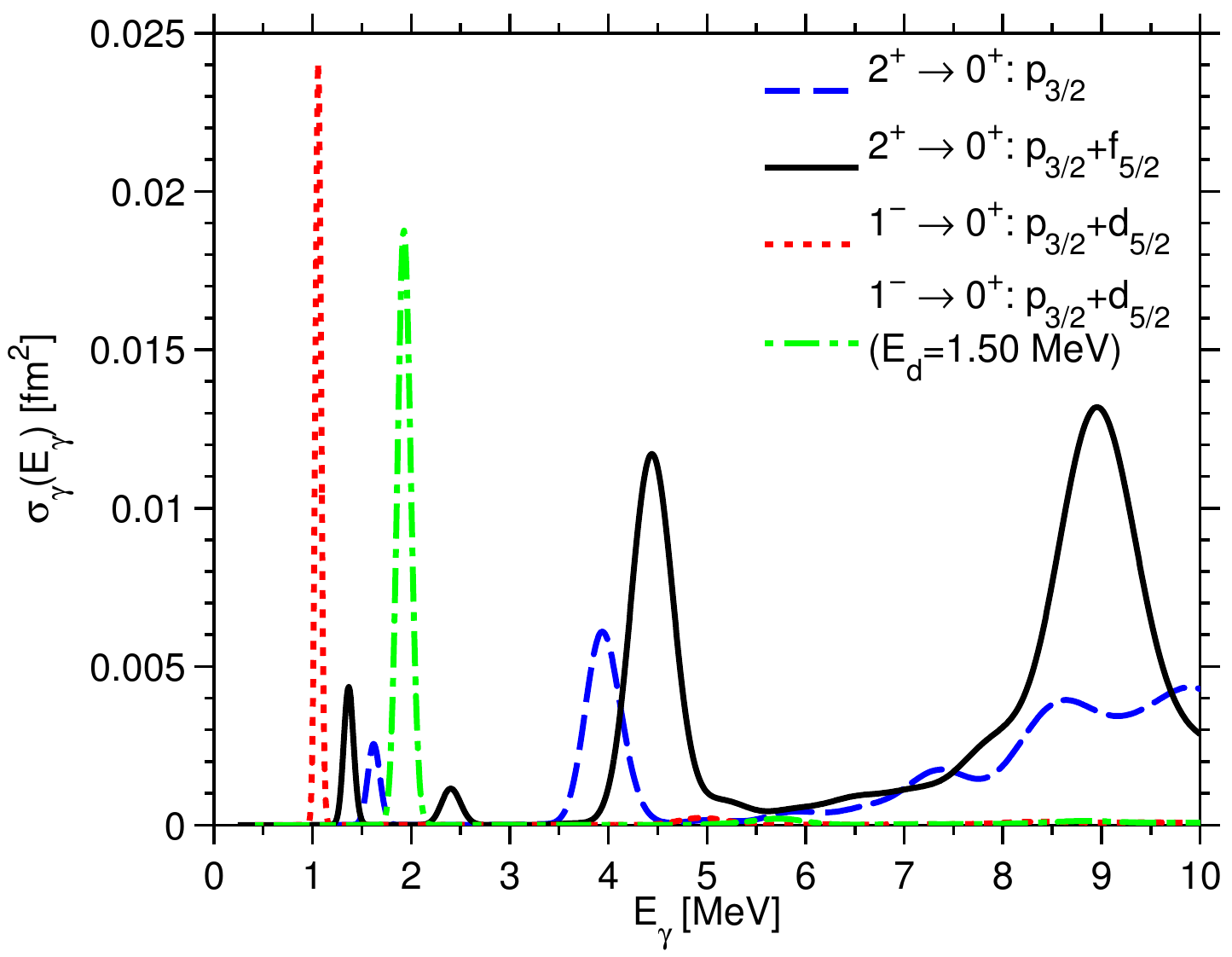}
  \caption{(Color online) The cross sections for $2^+ \rightarrow 0^+$ with both $p_{3/2}$ and $f_{5/2}$ waves (solid,
    black curve), and only with $p_{3/2}$ waves (dashed, blue curve). Also included are the cross section for $1^- \rightarrow 0^+$ with $p_{3/2}$ and $d_{5/2}$. For the red dotted curve $E_{d_{5/2}} = 0.641$ Mev, and $E_{d_{5/2}}=1.50$ MeV for the green dash-dotted curve. In both cases the $\mathcal{E}1$ transtition has been scaled down by $10^3$ to make the figure readable. The photodissociation energy is related to the total energy, $E$, and the binding energy, $B$, by $E_{\gamma} = E + |B|$. \label{fig cross}}
\end{figure}

In Fig.~\ref{fig cross} we show the cross section from Eq.~(\ref{eq
  siggam}) as function of energy above threshold for two cases of
$2^+$ and $1^-$ excitations each.  All the peaks are at the resonance energies where
the spatial overlaps to the ground state allow finite cross sections.
Two clear differences are seen between the two $2^+$ cases.  First of all,
the two lowest $p_{3/2} +f_{5/2}$ peaks are pushed a little bit further apart
as for a two-level system with an additional interaction. Secondly,
the peaks are notably smaller when only $p_{3/2}$ is allowed.  The
reason is the partial wave composition of the peaks and the number of
contributing potentials. 

All peaks are in the $p_{3/2}$ case composed entirely of $(l_x,l_y) =
(1,1)$, as seen in Table \ref{tab 3b excwave}, where only one potential contributes significantly to the
lowest peak.  In the $p_{3/2}+f_{5/2}$ case the same structure is
found in the lowest peak where two potentials now contribute evenly.
The next two peaks are in contrast composed of an almost even mixture
of $(l_x,l_y) = (1,3)$ and $(3,1)$, which clearly is not allowed with
only contribution from the $p_{3/2}$ state.  In between the resonance
peaks are the part of the cross section which makes the non-resonant
contribution to the reaction rate. This contribution is completely
negligible compared to the resonant contribution for low energies
while increasingly appearing at higher energies. As the lowest peak consists almost exclusively of p-waves, adding higher angular momentum orbitals would not change the cross section nor the resulting reaction rates.

The cross section for the $1^-$ excitation is also shown in Fig.~\ref{fig cross}. When the
$d_{5/2}$ and $p_{3/2}$ two-body states have the same energy we find
only one huge contribution at very low energy corresponding to one and
only one resonance. When the $d_{5/2}$ energy is increased this cross
section peak (resonance energy) moves up in energy, gets broader, and
decreases in size.  It is found that increasing the $d_{5/2}$ energy to 3.00 MeV makes the $\mathcal{E}1$ contributions
insignificantly small compared to the $\mathcal{E}2$ contributions in
Fig.~\ref{fig cross}.  We can conclude that $\mathcal{E}1$-transitions are large
even for relatively high-lying two-body states which are necessary for
the composition of the $1^-$ continuum or possible $1^-$ resonances.

\begin{figure}[t]
\centering
  \includegraphics[width=1\linewidth]{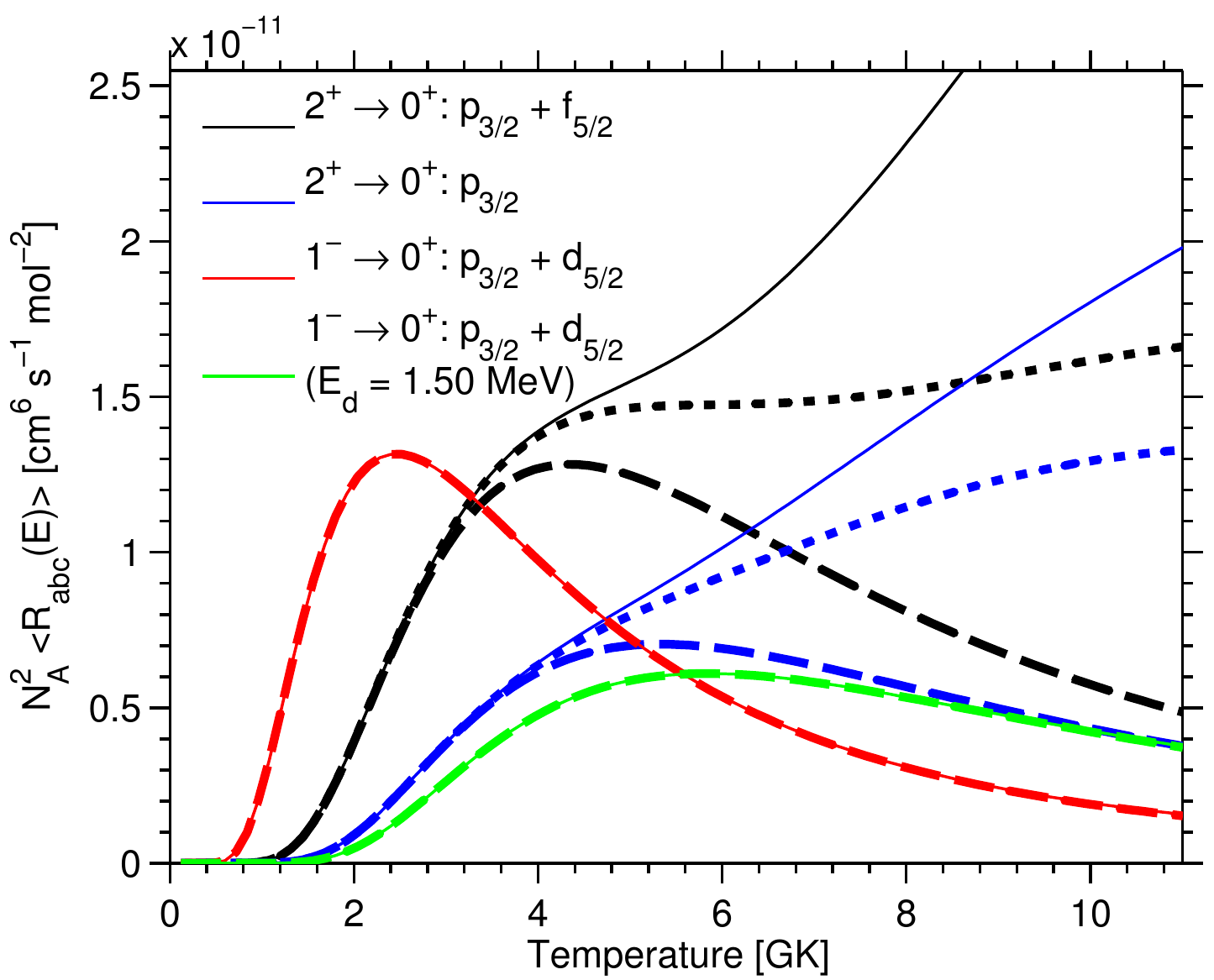}
  \caption{(Color online) The reaction rates corresponding to the cross sections from Fig.~\ref{fig cross}. The result of the full calculations is given by the solid curves. The dashed curves in the same colour are the result of applying Eq.~(\ref{eq rate esti}) to the lowest peak in Fig.~\ref{fig cross}. The dotted curves are the sum of contributions for the three (two) lowest resonances for $p_{3/2}+f_{5/2}$ $(p_{3/2})$ using Eq.~(\ref{eq rate esti}). The $\mathcal{E}1$ rates have been scaled down by $10^4$ to make the figure readable. The various resonance widths for the lowest resonances can be found in Table \ref{tab gamma3}. \label{fig rates}}
\end{figure}

The capture process takes place in an environment where temperature is an important parameter. The three-body energy is therefore not a priory given, but occurs with a certain probability distribution and with the capture rate specified in Eq.~(\ref{eq ave rate}).  The resulting rates are given by the full lines in Fig.~\ref{fig rates}. The lowest resonance peaks in Fig.~\ref{fig cross} suggest narrow states, which can be approximated by a Breit-Wigner shape. It is then possible to use the much simpler expression in Eq.~(\ref{eq rate esti}) to find the contribution from each resonance to the overall reaction rate.  

Figure \ref{fig rates} shows that the lowest resonance clearly dominates for both cases even at temperatures well above 3 GK. Summing the contribution from the isolated lowest resonances results in a better agreement with the full calculation to a higher temperature, but large deviations appear above around 5 GK. It is very remarkable that such a simple expression, based on a single resonance, is able to estimate the reaction rate so accurately over several GK.  It is even more impressive that a simple sum over the resonances can further increase the temperature range of its applicability while maintaining the accuracy. Increasing the $d_{5/2}$ energy increases the resonance energy, which in turn increases the temperature with highest rate, while simultaneously reducing this rate. The characteristic peak in the reaction curve is also smeared out by the effects of higher energy continuum contributions.

There are a number of possible corrections, which should be considered. First of all, the rates in Fig.~\ref{fig rates} are based on the full three-body $2^+$ ($1^-$) spectrum in the energy region seen in Fig.~\ref{fig cross}. If, as in the mirror nuclei \cite{tul04}, several of the lowest states are $2^+$ states, they are all included. Contributions from other possible states, such as $4^+$ or $3^-$ states, could conceivably contribute, but they would be suppressed by several orders of magnitude because of the higher order transition necessary. All transitions are to the same $0^+$ ground state. Excited, bound three-body states are not accounted for as they are very unlikely at the edge of the dripline, in particular considering the lowest excited state in the mirror nuclei is $0.94$ MeV above ground \cite{tul04}.

Due to the high temperatures involved core excitations could also be a contributing factor. The first excited state in $^{68}$Se is $E_{c1} = 0.854$ MeV above the ground state \cite{mcc12}. The probability of occupation at a temperature of 4 GK would then be $\exp\left( -E_{c1}/(k_BT) \right) = 0.09$. To get the reaction rate from Eq.~(\ref{eq ave rate}) we must multiply by $R_{cpp}$ which is proportional to the photodissociation cross section as parametrized in Eq.~(\ref{eq sig width}) for a resonance. The peak structure suggested by the parametrization is used to derive Eq.~(\ref{eq rate esti}) which assumes that the barrier penetration factor contained in $\Gamma$ and $\Gamma_{eff}$ has a relatively smooth energy dependence.  The numerical calculations of $\sigma_{\gamma}$ shown in Fig. \ref{fig cross} reveal the expected peak structure for small $E_{\gamma}$ confirming the sufficient smoothness of $\Gamma$ and $\Gamma_{eff}$. With everything being the same for ground and excited states we conclude that the contribution from low-lying excited states are reduced by their Boltzmann factor which then would be only a minor contribution. The second excited core state is $1.20$ MeV above ground, and is even less significant. To include contributions from core excitations or other bound three-body states one would have to include the partition functions of both the initial target and final three-body nucleus \cite{fow67,tho09}, to account for the thermal equilibrium between the ground and the excited states. This would in this case be minor corrections.

The discretized three-body states are limited to energies less than 10 MeV, so the rates must decrease for sufficiently high temperatures. Going to higher temperatures the computed $\mathcal{E}2$ transition rate has a maximum value of about 6.5 and 3.5 $\times \, 10^{-11} N_A^{2} \text{cm}^{6} \text{s}^{-1} \text{mol}^{-2}$ at just above 25 GK for the black and the blue curve respectively, and then decreases similarly to the $\mathcal{E}1$ transition. The maximum value for the $\mathcal{E}1$ transition occurs at much lower temperatures as there, as seen from Fig.~\ref{fig cross}, is no significant contribution to the cross section at higher energies. In both cases, the restriction imposed by the limit in discretized energies is only relevant far outside our region of interest.

As the single peak approximation fails at higher temperatures it would also fall short at lower temperatures. At low temperatures even the lowest resonance state would not have a significant probability of being populated, and the rate would be dominated by off-resonance contributions. Fortunately, this is well below our region of interest.

If both a $2^+$ and a $1^-$ state was present simultaneously transitions like $2^+ \rightarrow 1^- \rightarrow 0^+$ or $1^- \rightarrow 2^+ \rightarrow 0^+$ could be imagined, which would affect the final reaction rate. This effect could be computed by establishing the relations between the relevant two- and three-body energies, and then calculating the reaction rate in the same way as before. As before, the effect of any higher lying resonances would only be significant if the temperature was sufficiently high, as demonstrated by the accuracy of the lowest peak in isolation in Fig.~\ref{fig rates}. For our purposes it is therefore unnecessary to consider such corrections.

If a three-body force is added and the energy levels shifted, the effect on the cross section and reaction rate is only an energy scaling determined by Eqs.~(\ref{eq siggam}) and (\ref{eq rate esti}) respectively. The overlap matrix element would not change, as the energy shift does not change the structure of the wave function as seen from Table \ref{tab 3b wave}

\section{Practical implications \label{sec impli}}

The formulations and calculations presented in the previous sections
are schematic or quantitatively accurate depending on the point of
view.  The schematic impression arises from the relatively strong
assumptions of the few contributing proton-core partial waves.  This
is, however, not unrealistic within the field of few-body cluster
models, which has also been able to provide an appropriate description
of two weakly bound nucleons surrounding an ordinary nuclear core. 
The quantitative accuracy emerges as soon as the few-body model
approximations are shown to be correct, because then the reliability
and completeness of the results are unavoidable.

The final purpose of this paper is to provide a method for estimating proton capture reaction rates for the region around $A\sim 70$ and $N \sim Z$ in general, and the critical waiting points in particular. To that end we first discuss the reaction mechanism resulting from the sets of input parameters.  Second, we present estimates of the crucial two-and three-body ground and first excited state energies, and finally, we combine these considerations to present a general method for estimating radiative capture rates around the critical waiting points.

\subsection{Reaction mechanism \label{sec dir seq}}

The calculations are carried out without need for specification of the
reaction mechanism. In contrast, from the calculated results we can
deduce how the process proceeds between the borromean bound state and
the three free constituents in the continuum.  Due to the principle of
detailed balance both reaction directions are equally well suited for
both qualitative and quantitative descriptions.  In words the process
starts with bombarding the borromean bound state with photons of
energy larger than the three-body binding energy, where the populated
continuum states somehow end up as three independent infinitely
separated three particles, that is two protons and a core.

The process $A + \gamma \rightarrow p + c + p$ could proceed in several possible steps. It could include an excited continuum state of the borromean nucleus, $A^{\ast}$, or a quasi-stable two-body proton-core $(pc)$ configuration, although either or both of these steps might be skipped. In this way the process is divided in distinct, and possibly,
independent steps. Whether they are followed or not in a sequential progression is defining the reaction mechanism, and is used to categorize it. 

The traditionally denoted sequential path \cite{gri01,rod08} is with the proton-core
state, but with or without $A^{\ast}$,
\begin{alignat*}{3}
A + \gamma &\rightarrow A^{\ast} \rightarrow &(pc)+p &\rightarrow p+c+p, \\
A + \gamma &\rightarrow &(pc)+p &\rightarrow p+c+p. 
\end{alignat*}
Likewise, the route usually called direct \cite{gri05,rod08}, again with or without $A^{\ast}$, is
\begin{alignat*}{2}
A + \gamma &\rightarrow A^{\ast} \rightarrow &p+c+p, \\
A + \gamma &\rightarrow &p+c+p.
\end{alignat*}
The path chosen by nature depends on the characteristics of the system. However, substantial numerical simplifications as well as insight can be
gained with a dominating narrow resonance.  Approximating the cross
section by a peaked function like a Breit-Wigner shape we arrive at
the extremely simple temperature dependent rate expression in
Eq.~(\ref{eq rate esti}).  This is exactly the same expression as in
Eq.~(15) of Ref.~\cite{gar11}, with $\Gamma_{\gamma}$ replaced by
$\Gamma_{eff}$.  However, in Ref.~\cite{gar11} this reaction rate is
derived in the ``extreme sequential limit'', under the much stronger
conditions that $\Gamma_{\gamma}$ is much smaller than $\Gamma_{ppc}$,
and the direct decay is disallowed.  If $\Gamma_{\gamma} \ll
\Gamma_{ppc}$ then $\Gamma_{eff} \simeq \Gamma_{\gamma}$, and the
expressions become identical.  On the other hand, if $\Gamma_{ppc} \ll
\Gamma_{\gamma}$ then $\Gamma_{eff} \simeq \Gamma_{ppc}$, and photon
emission dominates over the strong decay channel.

Independent of validity of the simplified rate expression in
Eq.~(\ref{eq rate esti}), the process can still be either direct or
sequential, or for that matter any mixture.  This is perhaps better
appreciated by explicitly explaining that $\Gamma_{\gamma}$ is
determined entirely by the excited and ground state short-distance
properties, while the strong decay is entirely determined by the
structure underlying the barriers which has to be overcome before the
reaction is completed. The effective thickness of the barrier is determined by the three-body resonance energy level.

\begin{figure}[t]
\centering
  \includegraphics[width=1\linewidth]{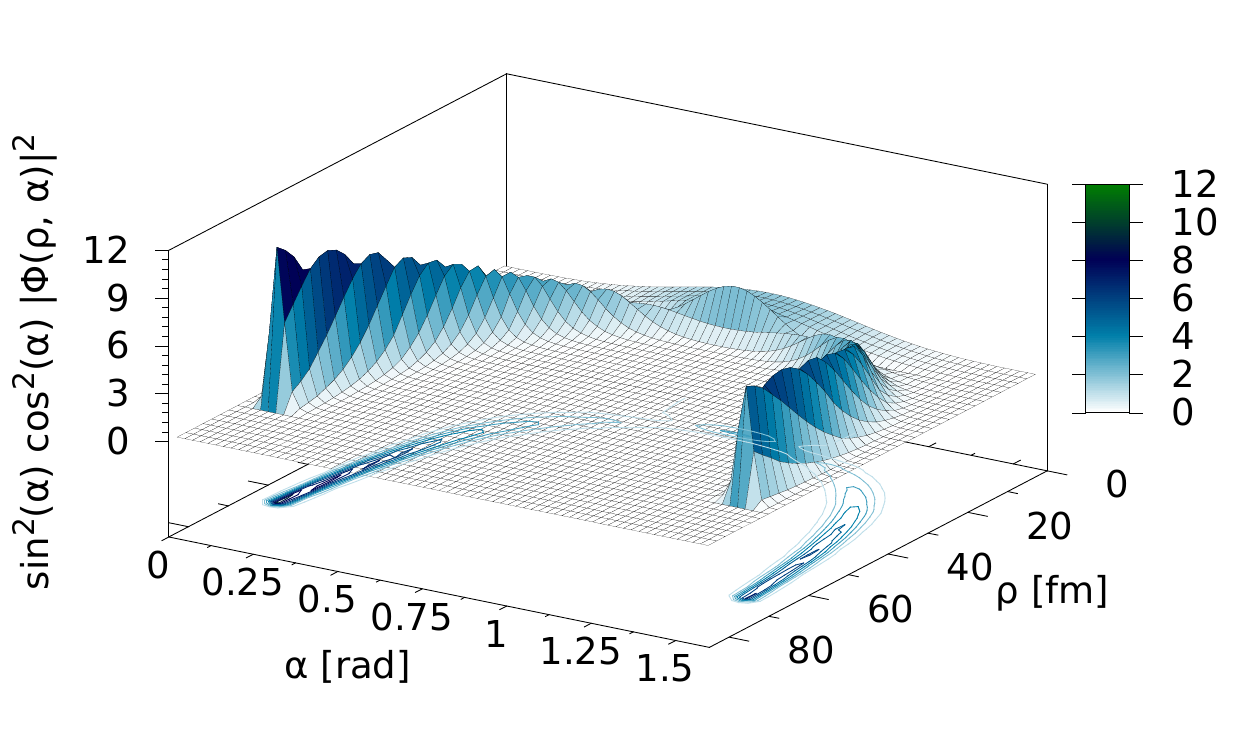}
  \caption{(Color online) The square of the angular wavefunction $\Phi$ from Eq.~(\ref{eq lamb}), multiplied by the phase factor $\cos^2(\alpha)\sin^2(\alpha)$ and integrated over $\Omega_x$ and $\Omega_y$, as a function of $\rho$ and $\alpha$ for the lowest $\lambda$ of the $2^+$ state in the second set of Jacobi coordinates.  \label{fig ang wave}}
\end{figure}

The intermediate structure can be directly investigated by the density distribution of the corresponding angular wave function from Eq.~(\ref{eq ang wave}). The angular wave function corresponding to the lowest $\lambda_n$ is shown in Fig.~\ref{fig ang wave} as a function of $\rho$ and $\alpha_{pc}$. The structure is abundantly clear for $\rho$ values larger than about $20$~fm. The probability distribution is only finite at large and small $\alpha_{pc}$, which through Eq.~(\ref{coord}) implies that the distance is small between either the core and the first proton, or the second proton and the centre of mass of the core-proton system. This is precisely the proton-core resonance configuration properly antisymmetrized. As this is the configuration for the lowest $\lambda_n$, the energetically most advantageous escape route is apparently through this configuration corresponding to sequential decay. However, this conclusion may change with relative sizes of the two- and three-body resonance energies, and as function of the total three-body energy. The angular wave functions corresponding to higher lying $\lambda_n$ eigenvalues have different configurations. By including all the relevant, relatively low-lying $\lambda$ spectrum, all the relevant angular configurations, and thereby all the relevant reaction mechanisms, are included, and not just the extreme sequential or direct paths.

The decay width from resonance to three particles can be estimated by use of the WKB tunnelling probability through the lowest potential barrier. This is most accurately done using numerical integration between the classical turning points where the effective potential equals the three-body resonance energy, $V_{eff}(\rho_0) = V_{eff}(\rho_t) = E_R$. Based on the energy difference from the potential minima to the resonance energy, the harmonic oscillator frequency $\omega_0$ can be determined. The knocking rate from Ref.~\cite{sie87} is about $\omega/\pi$ which must be multiplied by the second order WKB tunnelling probability, $1/(1+\exp(2S))$, that is
\begin{align} 
\Gamma_{ppc}  &=  \frac{\hbar \omega_0}{\pi}  \left( 1+ \exp(2S) \right)^{-1} \; , \label{wkb1} \\
S &= \frac{1}{\hbar} \int_{\rho_0}^{\rho_t} \sqrt{2 \mu_{p,cp} (V_{eff}(\rho)-E_R)} \; d\rho \; .  \nonumber
\end{align}

The $\Gamma_{\gamma}$ width can be calculated using Eq.~(\ref{eq gamgam}). The effective width $\Gamma_{eff}$ can be calculated from $\Gamma_{\gamma}$ and $\Gamma_{ppc}$. The values of these key quantities for the three-body decay, relating to the cases of
practical interest studied in the previous sections, are presented in Table \ref{tab gamma3}. We find in
almost all our borromean cases that $A^{\ast}$ is a well defined
resonance with a very narrow width. The reason is
readily found in the potentials, where the attraction is sufficiently
strong, inside very pronounced confining barriers, to hold a narrow
resonance. As seen in Eq.~(\ref{eq gamgam}) $\Gamma_{\gamma}$ depends on the resonance energy, but it is dominated by the overlap matrix element for the given multipole transition. As such it increases slowly with energy irrespective of partial waves, but it changes dramatically when the order of the transition is changed. The increase is much more drastic for $\Gamma_{ppc}$ as it depends exponentially on barrier thickness. 

\begin{table}[t]
\centering
\caption{The three-body resonance energy along with both photon decay width calculated using Eq.~(\ref{eq gamgam}) and proton decay width based on a WKB calcualtion. The effective width $\Gamma_{eff}$ is calcualted from these values. The cases of interest correspond to the peaks in Fig.~\ref{fig cross}. All energies and widths are in MeV. \label{tab gamma3}}
\begin{ruledtabular}
\begin{tabular}{l *{6}{l}}
State                  & $E_R$ & $\Gamma_{ppc}$       & $\Gamma_{\gamma}$    & $\Gamma_{eff}$       \\
\colrule
$2^+: p_{\frac{3}{2}}+f_{\frac{5}{2}}$ & 1.14  & $4.1 \cdot 10^{-9}$  & $5.3 \cdot 10^{-10}$ & $4.7 \cdot 10^{-10}$ \\
                       & 2.17  & $5.9 \cdot 10^{-3}$  & $8.3 \cdot 10^{-10}$ & $8.3 \cdot 10^{-10}$ \\
                       & 4.29  & $3.0 \cdot 10^{1}$   & $2.7 \cdot 10^{-8}$  & $2.7 \cdot 10^{-8}$  \\
$2^+: p_{\frac{3}{2}}$         & 1.38  & $2.4 \cdot 10^{-7}$  & $5.4 \cdot 10^{-10}$ & $5.4 \cdot 10^{-10}$ \\
                       & 3.68  & $9.6 \cdot 10^{-1}$  & $1.5 \cdot 10^{-8}$  & $1.5 \cdot 10^{-8}$  \\
$1^-: p_{\frac{3}{2}}+d_{\frac{5}{2}}$ & 0.61\footnote{$E_{2b}(p_{3/2}) = E_{2b}(d_{5/2}) = 0.641$ MeV.} &\multicolumn{1}{c}{--}& $1.7 \cdot 10^{-6}$  &\multicolumn{1}{c}{--}\\
                       & 1.50\footnote{$E_{2b}(p_{3/2}) = 0.641$ MeV, $E_{2b}(d_{5/2}) = 1.50$ MeV.}  & $5.2 \cdot 10^{-7}$  & $1.0 \cdot 10^{-5}$  & $4.9 \cdot 10^{-7}$ \\
\end{tabular}
\end{ruledtabular}
\end{table}

When the energy of the $1^-$ state is increased, by increasing the energy of the $d_{5/2}$
single-particle proton levels, the resonance becomes wider. The attractive pockets of the
potentials diminish while moving to higher energies, implying that the
resonance features disappear. The continuum states would then
have substantially less relative probability at distances comparable to the
size of the ground state. This decreases the rate of the $1^- \rightarrow 0^+$ transition. We emphasize that the method of the numerical calculations is completely unchanged, and the results are in
fact obtained without any information about whether such a resonance
exists or not.  

For the lowest resonance the $\Gamma_{ppc}$ is only about an order of magnitude larger than $\Gamma_{\gamma}$. This implies the lowest resonance is at the edge where the effective width becomes a mixture of the two. As $\Gamma_{ppc}$ decreases exponentially with $E_R$ without lower limit, while $\Gamma_{\gamma}$ decreases as $E_{\gamma}^{2\ell+1}$, where the photon energy necessarily is finite, $\Gamma_{eff}$ will be dominated by $\Gamma_{ppc}$ if the resonance energy is lowered.

If the three-body resonances, $E_R$, are the only intermediate
configurations acting as doorway states, then non-vanishing
processes occur only when $E \approx E_R$ where the uncertainty is
determined by the width of the states.  On the other hand, small rates
and cross sections arise also for energies without the match to
resonance doorway states.  The rate is calculated from the theoretical
formulation for given energy, but the applications are rather for given
temperature, $T$. The temperature smearing over the energies up to
around $T$ then all contribute, and the higher energy contributions
become exponentially suppressed. A much more detailed discussion can be
found in Ref.~\cite{gar11}.

The character of the reaction mechanism as sequential, direct, a
mixture or something else, is fundamentally determined by the dynamic
evolution from small to large distances.  However, a number of rather
solid conjectures can be made from the calculated static properties.
This can be elucidated with two schematic potentials based on extreme
geometric progression corresponding to sequential and direct decay, for more details see Ref.~\cite{gar04c}.
Outside the strong attractive region the Coulomb potential for one
proton moving away from proton-core state of given positive energy,
$E_{pc}$, is given by $V_{seq}=(Z_c+1)e^2/\rho + E_{pc}$. In case of several possible, different two-body energies the most likely potential to tunnel through would be the small and narrow potential. If the two
protons are moving away from the core along the most favoured symmetric
linear configuration with the same $\rho$ the potential is instead,
$V_{dir}=(2Z_c+1/2)e^2\sqrt{2}/\rho$, where $Z_c |e|$ is the core
charge.  These two potentials,  $V_{seq}$ and $V_{dir}$, cross each other at $\rho_c$ for an
energy $V_c$, that is
\begin{align} \label{eq cros1}
 \rho_c &= \frac{e^2 \left( (2\sqrt{2}-1)Z_c -1 + 1/\sqrt{2} \right)}{E_{pc}} \approx \frac{2.6 Z_c}{E_{pc}} \;,\\ 
 V_c &=  E_{pc}\frac{(2Z_c+1/2)\sqrt{2}}{(2\sqrt{2}-1)Z_c -1 + 1/\sqrt{2}} \approx  1.6 E_{pc} \; \label{eq cros2}
\end{align}
for $Z_c \gg 1$.  With $E_{pc} = 0.641$~MeV and $Z_c =34$, we get
$\rho_c=139$~fm and $V_c = 1.00$~MeV, which is significantly further out than the potential thickness in Fig.~\ref{fig pot 2+}a. The most energetically favourable decay path is then the sequential path, barring drastic, and most likely energy demanding, changes in the configuration during the decay process. As $E_{3b} \simeq E_{2b}$ for $1^-$ the potential thickness is tending towards infinity, which would exclude a sequential decay.

The reaction mechanism for photo dissociation, or equivalently
radiative capture, of given energy $E$ can then be expected
characterized as sequential for $ V_c < E$, mixed sequential direct for
$ E_{pc} < E < V_c$, mixed direct and virtual sequential $E < E_{pc}
< V_c $, and direct $ E << E_{pc}$ \cite{rod08,jen10}.  We emphasize that these
characteristics are not rigorous properties, although limiting cases
would be observable \cite{che07}, as they would leave distinct signatures in the energy
distribution of the emerging fragments. Based on these limits the reaction mechanism can very convincingly be classified as sequential for the $2^+$ case, direct for the low-lying $1^-$ case, and either sequential or a mixture for the high-lying $1^-$ case. It should be noted that this is very strongly dependent on the specific potential depths and the temperature in question. For temperatures significantly higher or lower than the resonance energy the reaction mechanism would not be determined by the single lowest resonance, but by the continuum background contribution or possibly a complicated combination of several different resonances.

\subsection{Energy level predictions \label{sec engy lim}}

The central parameters in the effective lifetimes of the critical waiting points are the proton binding energies of the two following isotones. Ideally, one would like to predict these two- and three-body energies exactly. Unfortunately, insufficient experimental knowledge makes such predictions difficult. However, as seen in Sec.~\ref{sec engy}, it is possible to establish relations between the possible two-body energies and the needed three-body energies. From there limits can be inferred regarding the position of low lying excited energy levels in both the two- and three-body system.

As the critical waiting points are borromean in nature the core-proton system must be unbound, while the two-proton system must be bound. In other words, the region of interest is limited to $E_{2b}>0$ and $E_{3b}<0$. The following limits are calculated without including a three-body potential. Adding an attractive three-body potential would lower the three-body energy, and thereby increase the estimated energy ranges.

If the ground state proton separation energy, $S_p$, is greater than 0.35 MeV (where the highest single wave curve in Fig.~\ref{fig ener} crosses zero), then there must be a very close lying first excited core-proton state for the three-body system to be bound. Depending on the $S_p$ value a very narrow energy range for this excited level can be predicted. Likewise, $S_p$ cannot be greater than 0.74 MeV (where the $p_{3/2}+f_{5/2}$ curve crosses zero in Fig.~\ref{fig ener}), as the three-body system then could not be bound. Of course, this is based on the assumption that shell model predictions and the mirror nuclei correctly identifies the relevant single-particle states in the given region.

The only thing missing, if proper estimates are to be made concerning the waiting point nuclei, is the two-body energies. However, even this is not as severe a limitation as might be imagined. Based on the results in Fig.~\ref{fig ener} energy intervals can be established for the proton separation energy. Recently, it was possible to measure the ground state proton separation energy of $^{69}\text{Br}$ to $S_p(^{69}\text{Br}) = 641(42)$~keV \cite{san14}. For this to comply with our result the ground state must be either a $p_{3/2}$ or a $f_{5/2}$ state, with the other being a low lying excited state, otherwise the three-body system would be unbound. In Ref.~\cite{san14} the ground state is surmised to be a $f_{5/2}$ state with a $p_{3/2}$ state lying an unknown distance above. All lines allowing only one partial wave have crossed into the unbound three-body region before $E_{2b} = 0.641$~keV. The same is true for the $f_{5/2}+p_{1/2}$ line. The only remaining possibility among the likely partial waves is the $p_{3/2}+f_{5/2}$ combination.

The upper and lower limits for this first excited state can now be established from the relation between two- and three-body energy. The lower limit is of course $E_{2b}(p_{3/2}) = 0.641$~MeV, corresponding to a degenerate ground state. The upper limit is determined by keeping the $f_{5/2}$ energy constant and varying the $p_{3/2}$ energy. Setting $E_{2b}(f_{5/2}) = E_{2b}(p_{3/2}) = 0.641$~MeV results in $E_{3b}=-0.21$~MeV. By slowly changing $E_{2b}(p_{3/2}$) the upper limit is found to be at $E_{2b}(p_{3/2}) = 0.800$~MeV, where $E_{3b}=-0.00$~MeV. The first excited $p_{3/2}$ state in $^{69}\text{Br}$ is then predicted to lie less than $0.16$~MeV above the ground state. Likewise, using the estimates included in the AME 2012 collection, a proton separation energy of $S_p(^{73}\text{Rb}) = 0.6$~MeV is predicted \cite{aud12}. This implies the ground and first excited states consist of a $p_{3/2}$ and $f_{5/2}$ state, one lying no more than 0.2 MeV from the other. For $^{65}\text{As}$ the AME estimated proton separation energy is at $0.09$ MeV, which makes it difficult to predict anything specifically. However, it does make it very unlikely that there is a $p_{1/2}$ or a $f_{5/2}$ ground state without a low lying first excited state.

Similar predictions can be made concerning the excited continuum levels. Based on Fig.~\ref{fig engy cont} the $2^+$ state is found to very consistently be $1.38$ MeV above the ground state. The lower limit of the resonance energy, $E_R$, then corresponds to the lower limit of the ground state, i.e. $E_R^{\text{min}} = 2 \times 0.641 \, \text{MeV} - 1.54 \,  \text{MeV} + 1.38 \, \text{MeV} = 1.12 \, \text{MeV}$. Likewise, the upper limit corresponds to where the ground state is at the edge of being unbound, i.e. $E_{2b}=0.74$ MeV and $E_R^{\text{max}} = 1.38$ MeV. This very narrow interval can be used to estimate limits for the reaction rate.

It is considered much more unlikely that the three-body system will form a $1^-$ state based on both mean field calculations and comparisons with mirror nuclei. However, as has been shown, if it is even remotely possible the dipole transition will dominate the reaction rate. The placement of the $1^-$ resonance is not as sharply limited as the $2^+$ resonance, which is dictated by the lowest two-body resonance. On the other hand, the $1^-$ resonance needs a combination of opposite parity states, and could move to arbitrarily high energies without affecting the $0^+$ ground state.

\subsection{Rate predictions \label{sec rates}}

In the preceding sections it was argued that very narrow resonances will be produced by the fairly attractive short-distance region in combination with the wide Coulomb barrier at large distances. In addition, very confining limits have been placed on both two- and three-body energy levels. Collectively, this allows for limits to be placed on the relevant three-body reaction rates for specific energies. These limits would again be affected by the addition of a three-body potential. However, the energy scaling is predicted by Eqs.~(\ref{eq siggam}) and (\ref{eq rate esti}). More generally, this provides a method for estimating proton capture rates for the region in general given few experimental data.

As the reaction rate is dominated by narrow resonances it can well be approximated by Eq.~(\ref{eq rate esti}) for temperatures in the $0.1 - 4$ GK range. The limits of the resonance energy was established in Sec.~\ref{sec engy lim} for the specific case of the critical waiting points. Assuming the overlap matrix element is constant for small changes in two-body energy, $\Gamma_{\gamma}$ only depends on photon energy as $E_{\gamma}^{2\ell+1}$. However, as the distance between the curves in Fig.~\ref{fig engy cont} is constant, $E_{\gamma}$ would also be constant independent of two-body energy. The same is therefore true of $\Gamma_{\gamma}$, and the $\Gamma_{\gamma}$ value from Table \ref{tab gamma3} can be used for a given resonance. As $\Gamma_{eff}^{-1} = \Gamma_{ppc}^{-1} + \Gamma_{\gamma}^{-1}$ the largest possible value of $\Gamma_{eff}$ is the constant $\Gamma_{\gamma}$.

It is then clear that using Eq.~(\ref{eq rate esti}) the largest rate is achieved at the lowest resonance energy where $\Gamma_{eff} \simeq \Gamma_{\gamma}$. At some point, for sufficiently low resonance energies, the $\Gamma_{eff} = \Gamma_{\gamma}$ assumption no longer holds as $\Gamma_{ppc}$ continues to decrease. For lower resonance energies $\Gamma_{eff} < \Gamma_{\gamma}$ and the rate decreases.

To estimate the rate it is then necessary to estimate $\Gamma_{ppc}$ which means the barrier must be determined very accurately. Unfortunately, the intermediate distances that are relevant here are notoriously difficult to treat accurately by simple expressions. For large distances approximation such as the $V_{dir}$ and $V_{seq}$ potentials presented earlier could be used. These will however overestimate the potential, which will underestimate of the rate exponentially. A more appealing alternative is to use the potential calculated in full in Sec.~\ref{sec cont}. The potential outside the barrier is the determining part. To study the rate based on a two-body energy different from the value of $0.641$ MeV this part of the potential needs to be shifted an amount corresponding to the difference in two-body energies.

The final procedure in estimating the three-body reaction rate is as follows. The first step is to asses the two-body energies somehow. For waiting point nuclei limits can be established from Fig.~\ref{fig ener}. The second step is to find the three-body energies based on the two-body energies. The ground state three-body energy is determined from Fig.~\ref{fig ener}, while the first excited level is determined from Fig.~\ref{fig engy cont}. If different two-body energies are used for the relevant partial waves the lower limit for the three-body energy is given by the curve where the energy is the same, while the upper limit is given by the highest single wave curve. The third step is to estimate $\Gamma_{eff}$. The value of $\Gamma_{\gamma}$ is given in Table \ref{tab gamma3} for the most likely cases. Otherwise the single-particle decay rates can be used to estimate $\Gamma_{\gamma}$ \cite{sie87}, where a factor of two should be added to account for the two particles. In many cases $\Gamma_{\gamma} \ll \Gamma_{ppc}$, and the exact value of $\Gamma_{ppc}$ is then not relevant. When necessary, for relatively low resonance energies, $\Gamma_{ppc}$ can be estimated with a WKB calculation through the known potential barrier shifted appropriately according to the two-body energy. The fourth and final step is to calculate the limits of the reaction rate with Eq.~(\ref{eq rate esti}) using the calculated effective width and the limits on the resonance energy. This method applies generally for proton capture in the region around the critical waiting points. If considering $^{64}$Ge or $^{70}$Kr instead of $^{68}$Se minor changes in the long-range Coulomb potential would have to be considered, but the most important change would be the two- or three-body energy spectrum. The same method could be applied based on estimates of the particular energy levels.

\begin{figure}[t]
\centering
  \includegraphics[width=1\linewidth]{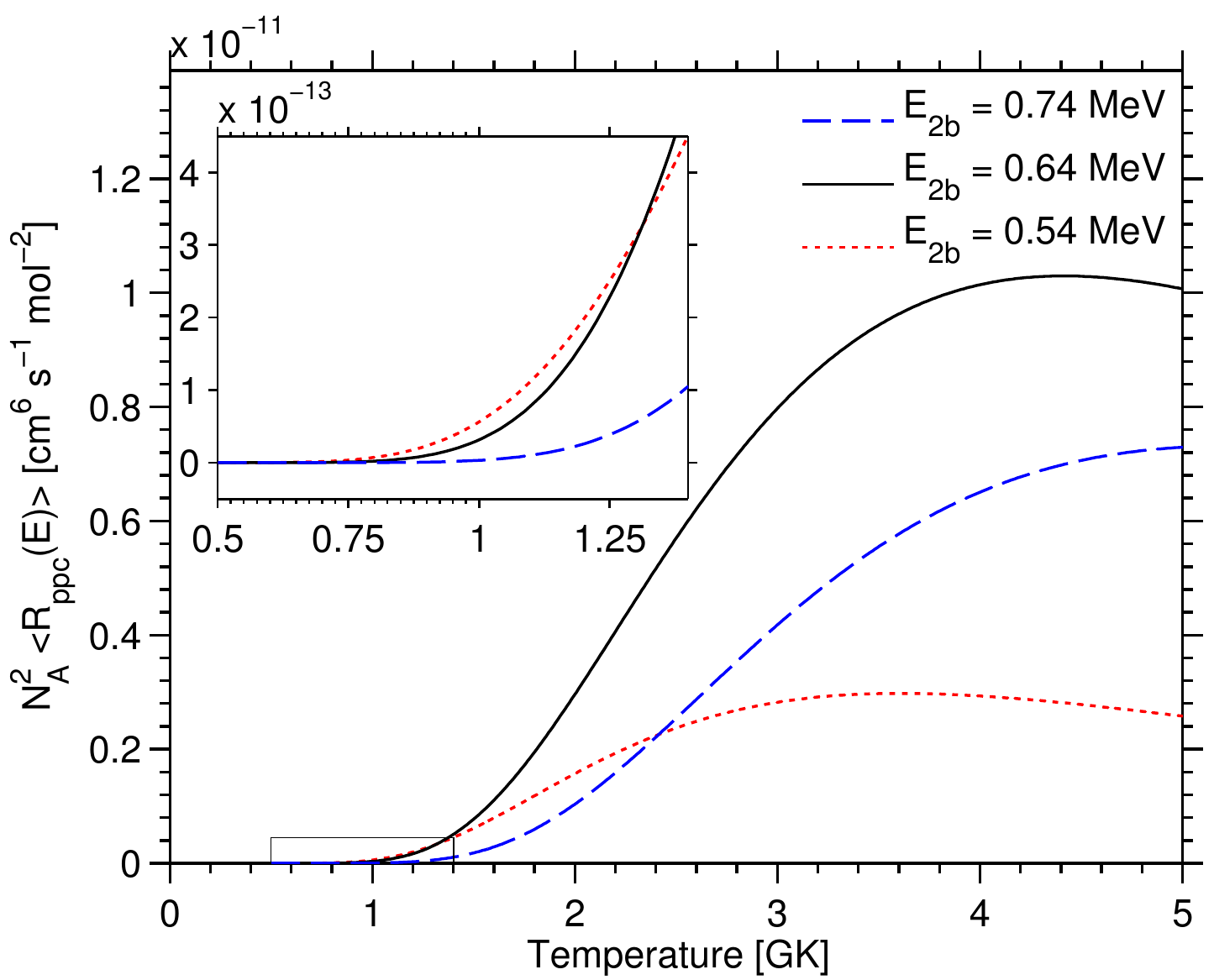}
  \caption{(Color online) Estimates of the reaction rate for $2^+ \rightarrow 0^+$ with $p_{3/2}$ and $f_{5/2}$ based only on two-body energies. The blue curve at $E_{2b}=0.74$ MeV corresponds to the upper limit established in Sec.~\ref{sec engy lim}. The low temperature region has been scaled up to demonstrate the similarities. \label{fig lim rate}}
\end{figure}

The result is seen in Fig.~\ref{fig lim rate}, where the $2^+ \rightarrow 0^+$ rate is estimated based on three different two-body energies. The energy studied in Sec.~\ref{sec cont} is used along with the upper energy level of $E_{2b}=0.74$ MeV established in Sec.~\ref{sec engy lim} and an energy an equal amount below. The low temperature region is scaled up to show how closely the rates agree for temperatures below $\sim 1$ GK. The rate decreases with increasing resonance energy because of the exponential factor in Eq.~(\ref{eq rate esti}), and it also decreases with decreasing resonance energy because we are at the edge where $\Gamma_{ppc}$ starts affecting $\Gamma_{eff}$. Changing the various two-body energies individually, between $0.54 - 0.74$ MeV, would change the three-body resonance energy, but the final reaction rate would be within the limits here presented. Going outside this two-body energy region would decrease the rate exponentially, based on Eq.~(\ref{eq rate esti}), for temperatures around 1-4 GK.

These limits are based on the assumption that only the lowest resonance contributes significantly. The temperature must be in the vicinity of the three-body energy for this to be true. For much lower temperatures the resonance energy is not accessible, and the main contribution is through off-resonant reactions. This would lower the rate exponentially. For higher temperatures other resonances would contribute, and the lowest resonance might only contribute through the tail of the cross section.

In principle, these rate estimates should include the various contributions discussed at the end of Sec.~\ref{sec reac res}, even for temperatures in the 1-4 GK region. However, these are all minor corrections which does not change the result significantly. Major changes would only occur with changes in energy levels.

\section{Conclusion \label{sec con}}

The waiting points in the rp process are prime candidates for three-body calculations by their very definition. Given the heavy core in the three-body system the proton-core interaction will determine the characteristic of the system, with only minor corrections stemming from the proton-proton interaction. Because of this, very simple relations between the two- and three-body energies can be derived and understood based on the three-body Hamiltonian. These relations apply equally well for both ground and excited energy levels. These very simple, and rather general relations between two- and three-body structures and energies allow for the estimation of either two- or three-body properties given very sparse experimental data.

It is also seen that the three-body structure consists of the two  protons located at the surface of the heavy core. At larger distances the structure is that of single proton moving away from a two-body core-proton resonance. Conceptually, this corresponds to what is traditionally known as sequential decay. For low available initial three-body energy (and low temperature) this configuration would no longer be the energetically most favourable.

When calculating the cross section for the process $A + \gamma \rightarrow p+p+c$ the effective potential confining the particles is decisive. The relatively heavy core gives rise to a large Coulomb barrier and an attractive short-distance potential. This makes the lowest continuum states narrow, well-defined resonances. Based on the spectrum of the mirror nuclei, $2^+$ states were assumed to dominate the low-lying spectrum. The full three-body rate calculation was based on the entire $2^+$ spectrum. Assuming the narrow, well-defined resonances have a Breit-Wigner shape, the full rate calculation could be reproduced from around 0.5 GK and up to 4 GK based solely on the lowest resonance. Summing over the contributions from the first few resonances for the $2^+$ case increased this temperature range to about $5-6$ GK. The off-resonance, background contribution is mainly relevant for lower temperatures, where the chance of accessing the resonance level is much smaller. For the much more unlikely $1^-$ case only one resonance state is present, which then dominates a wide temperature range around the resonance energy. 

In the unlikely scenario where an $1^-$ state is available, depending on its position, the $\mathcal{E}1$ transition could contribute significantly. However, based on both mean field calculations and experimental measurements in the region around the critical waiting points single-particle orbitals of like parity dominate the low-lying energy levels. Also no $1^-$ state is seen in the mirror nuclei. This very effectively excludes three-body states of negative parity. The most likely transition is then the always allowed (as long as more than s-waves are available) $\mathcal{E}2$ transition.

As the expression based on the assumed Breit-Wigner shape depends mainly on the resonance energies, the relations between two- and three-body energies can be used to provide limits for the reaction rate. Alternatively, the rate estimates could also be based directly on three-body resonance energies if available. 

An added benefit of the three-body formalism is that no assumptions regarding the preferred reaction path are needed. On the contrary, the most likely reaction mechanism can be deduced based on the structure of the angular wave functions corresponding to the lowest $\lambda_n$ eigenvalues. For most of the relevant energies the mechanism can be considered sequential through narrow two-body resonances. This is intuitively understandable if the three-body excited resonance state has a higher energy than the resonance(s) of the subsystem.  On the other hand, a very low temperature only allows direct decay (or capture) since then the available energy is too low to populate the two-body resonances even virtually.  However, these reaction questions would eventually have to be answered by studying the dynamic evolution from small to the large distances in more detail.

In conclusion we have performed a full three-body analysis of the nuclear structure and decay of critical waiting points in the rp process. This allowed us to study in detail the two-proton capture rate needed to bridge the waiting points, as well as the energy levels central in determining the effective lifetimes of these waiting points in a stellar environment. We find that a simple expression, based on an assumed Breit-Wigner shape, can accurately reproduce the full three-body rate calculation in the temperature region $0.5-5$ GK generally considered to be of astrophysical interest. This led to a general method for estimating two-proton capture rates in the region around the critical waiting points based only on either the two- or three-body energies, if these are more readily available. Specifically, we predict, given the currently available experimental data, that the two-proton capture rate forming $^{70}\text{Kr}$ for temperatures between 2 and 4 GK  increase by a factor of 3.5 from $0.4 \cdot 10^{-11} N_A^{2} \text{cm}^{6} \text{s}^{-1} \text{mol}^{-2}$.

\begin{acknowledgments}

The authors are grateful to K. Riisager for fruitful discussions and thoughtful insights. This work was funded by the Danish Council for Independent Research DFF Natural Science and the DFF Sapere Aude program, and partly supported by funds provided by DGI ofMINECO(Spain) under contract No. FIS2011-23565. We also acknowledge financial support from the European Research Council under ERC starting grant “LOBENA”, No. 307447.

\end{acknowledgments}

\end{document}